\documentclass[review]{elsarticle}

\usepackage{lineno,hyperref}
\usepackage[utf8x]{inputenc}
\usepackage{lmodern,textcomp}
\usepackage{amsmath}
\modulolinenumbers[5]

\journal{JQSRT}

\begin{document}

\begin{frontmatter}

\title{Measuring night sky brightness:\\ methods and challenges}

\author{Andreas H\"anel$^{1}$, Thomas Posch$^{2}$, Salvador J. Ribas$^{3,4}$, Martin Aubé$^{5}$, Dan Duriscoe$^{6}$, Andreas Jechow$^{7,13}$, Zoltán Kollath$^{8}$, Dorien E. Lolkema$^{9}$, Chadwick Moore$^{6}$, Norbert Schmidt$^{10}$, Henk Spoelstra$^{11}$, Günther Wuchterl$^{12}$, and Christopher C. M. Kyba$^{13,7}$}
\address{$^{1}$Planetarium Osnabrück, Klaus-Strick-Weg 10, D-49082 Osnabrück, Germany\\ 
$^{2}$Universität Wien, Institut für Astrophysik, Türkenschanzstraße 17, 1180 Wien, Austria
tel: +43 1 4277 53800, e-mail: thomas.posch@univie.ac.at (corresponding author)\\
$^{3}$Parc Astronòmic Montsec, Comarcal de la Noguera, Pg. Angel Guimerà 28-30, 25600 Balaguer, Lleida, Spain\\
$^{4}$Institut de Ciències del Cosmos (ICCUB), Universitat de Barcelona, C.Martí i Franqués 1, 08028 Barcelona, Spain\\
$^{5}$D\'{e}partement de physique, C\'{e}gep de Sherbrooke, Sherbrooke, Qu\'{e}bec,  J1E 4K1, Canada \\
$^{6}$Formerly with US National Park Service, Natural Sounds \& Night Skies Division, 1201 Oakridge Dr, Suite 100, Fort Collins, CO 80525, USA\\
$^{7}$Leibniz-Institute of Freshwater Ecology and Inland Fisheries, 12587 Berlin, Germany\\
$^{8}$Eötvös Loránd University, Savaria Department of Physics, Károlyi Gáspár tér 4, 9700 Szombathely, Hungary\\
$^{9}$National Institute for Public Health and the Environment, 3720 Bilthoven, The Netherlands\\
$^{10}$DDQ Apps, Webservices, Project Management, Maastricht, The Netherlands\\
$^{11}$LightPollutionMonitoring.Net, Urb. Ve\"{i}nat Verneda 101 (Bustia 49), 17244 Cass\`{a} de la Selva, Girona, Spain\\
$^{12}$Kuffner-Sternwarte,Johann-Staud-Straße 10, A-1160 Wien, Austria\\
$^{13}$Deutsches GeoForschungsZentrum Potsdam, Telegrafenberg, 14473 Potsdam, Germany}


\begin{abstract}
Measuring the brightness of the night sky has become an increasingly important topic in recent years, as artificial lights and their scattering by the Earth’s atmosphere continue spreading around the globe. Several instruments and techniques have been developed for this task. We give an overview of these, and discuss their strengths and limitations. The different quantities that can and should be derived when measuring the night sky brightness are discussed, as well as the procedures that have been and still need to be defined in this context.
We conclude that in many situations, calibrated consumer digital cameras with fisheye lenses provide the best relation between ease-of-use and wealth of obtainable information on the night sky. While they do not obtain full spectral information, they are able to sample the complete sky in a period of minutes, with colour information in three bands. This is important, as given the current global changes in lamp spectra, changes in sky radiance observed only with single band devices may lead to incorrect conclusions regarding long term changes in sky brightness. The acquisition of all-sky information is desirable, as zenith-only information does not provide an adequate characterization of a site. 
Nevertheless, zenith-only single-band one-channel devices such as the “Sky Quality Meter” continue to be a viable option for long-term studies of night sky brightness and for studies conducted from a moving platform. Accurate interpretation of such data requires some understanding of the colour composition of the sky light. We recommend supplementing long-term time series derived with such devices with periodic all-sky sampling by a calibrated camera system and calibrated luxmeters or luminance meters.
\end{abstract}


\end{frontmatter}


\section{Introduction}
The last decade has seen a rapid increase in research into artificial light in the nighttime environment (Table 1). This increased attention can be attributed to several factors, including: recognition of the impacts of artificial light on ecology and health \cite{long1,HolkerWolter1,stevens2013adverse}, increasing amounts of artificial light in the environment \cite{holker2010,miguel2014}, improved quality of imagery from space \cite{rs5126717,Miguel01082014}, the current global change in lighting technology \cite{de2014solid}, and increasing quality, ease, instruments, and methods for measuring light at starlight intensities \cite{cinzano2005,kollath2010}.

\begin{table}[h]
\centering
\caption{Bi-yearly citations counts to the seminal "World Atlas" paper by Cinzano et al.\ \cite{Cinzano2001} (WA), the "Ecological Light Pollution" paper by Longcore \& Rich \cite{long1} (LR), or references to ``Sky Quality Meter'' (SQM) in Google Scholar, 2005--2016. Search performed on 10th January 2017.}
\label{my-labela}
\begin{tabular}{lllllll}
     & 2005--06 & 2007--08 & 2009--10 & 2011--12 & 2013--14 & 2015--16\\ \hline
WA   & 26       & 46       & 39       & 75       & 120      & 130 \\
LR   & 22       & 45       & 63       & 87       & 160      & 180 \\
SQM  & 8        & 24       & 36       & 72       & 89       & 110 \\
Sum  & 56       & 115      & 138      & 234      & 369      & 420
\end{tabular}
\end{table}
While the increase in possible ways to measure artificial light is a positive development, it likely presents a barrier to newcomers to the field of light measurement. Furthermore, both the various terms and units used in standard and astronomical photometry and the typical amounts of light experienced in the environment are not familiar to most scientists. This presents a challenge for interdisciplinary researchers, such as biologists who wish to measure the light exposure during a field experiment (see the discussion in \cite{swaddle2015framework} and examples of such studies \cite{pendoley2012, jechow2016}). 

This paper aims to introduce measurements of night sky brightness at visible wavelengths to readers with no background in the field. While the focus is the night sky, we expect this overview will be useful for readers interested in characterization of field sites for biological studies\footnote{Such readers should however keep in mind that properly characterizing sites for biological studies also involves measuring glare, and potentially ultraviolet and infrared radiation as well.}. We first provide a background on natural and artificial night sky brightness (Section 2). We then discuss different techniques for measuring sky brightness, starting with the visibility of stars and traditional naked-eye observations (Section 3), followed by measurements with single channel instruments (Section 4), imaging instruments (Section 5), and spectrometers (Section 6).  The classic technique of astronomical photometry is discussed in the Appendix, as it is unlikely to be used by researchers outside of the field of astronomy. Unless otherwise stated, the paper assumes that direct sources of light are not present in the field of view of the instrument.

It is important to note that in many cases we convert values into SI units for the sake of facilitating understanding, but in most cases this conversion is only approximate. For example, an inexpensive luxmeter that does not perfectly match the CIE spectral sensitivity \cite{CIE} could give reasonably accurate readings for a source with a daylight-like spectrum, but have large errors for sources with a different spectral power distribution (such as sodium or fluorescent lamps).

\section{Sky brightness}
\label{sec:sky_brightness}
Near human settlements, the brightness of the night sky is made up of light from both natural and artificial sources. Celestial light can travel directly to an observer (allowing one to see stars, galaxies, etc.), or it can scatter in the atmosphere and contribute to the diffuse glow of the sky. Some celestial sources appear as points or small objects (e.g.\ stars, planets, the moon), others are extended and diffuse (aurorae, airglow, galactic nebulae, galaxies). Even after setting, the sun can contribute diffuse light through scattering interactions with the atmosphere (twilight and noctilucent clouds) or from dust in the solar system (zodiacal light and Gegenschein). Details about many of these natural sources are discussed by Leinert et al.\ \cite{Leinert1} and Noll et al.\ \cite{noll2012}, and typical values for night sky brightness are presented in Table 2.

Artificial light emitted from the Earth’s surface can also scatter off molecules or aerosols in the atmosphere and return to Earth as ``skyglow'' \cite{chant1935, garstang1986, Aub20140117}. On clear nights, this brightened sky results in a loss of star visibility \cite{2001MNRAS}, especially near the horizon where the glow is brightest. It also reduces the degree of linear polarization of the clear moonlit sky \cite{kyba2011}. Clouds have a strong influence on both natural and artificial sky brightness. In urban areas, clouds can increase skyglow by more than an order of magnitude, whereas in natural areas devoid of artificial light clouds make the sky modestly darker \cite{kyba2015,phdthesis,ribas2016}.

There are two parameters that can be used to quantify light in the environment. The first is “irradiance”, which can be thought of as the total amount of electromagnetic radiation falling on a surface. A well-lit office likely has a high level of irradiance, and relatively uniform values across over the floor and surfaces in the room. The second is “radiance”, which can be thought of as how bright a (usually small) area in your field of view appears. The unpleasant sensation felt when looking directly towards a glaring lamp is due to the large radiance. Spectrometers measure radiance (or irradiance) at many different wavelengths. The radiance per unit wavelength is called the spectral radiance. Other devices have a more restricted range. In this case, the spectral sensitivity of the device must be reported along with the radiance value. Photometric instruments are specially designed so that their spectral response matches a simplified theoretical response curve for the human visual system \cite{CIE}. When the spectral response is perfectly matched to human vision, then measurements from these instruments are called “luminance” (or popular "brightness") and “illuminance”, instead of “radiance” and “irradiance”. For measurements of the sky brightness, it is important to further clarify whether sky radiance is being measured in the areas between the visible stars (``sky background brightness''), or whether starlight is included (``sky brightness''). Irradiance always includes all sources of light.

Traditionally, sky brightness is measured by astronomers in the astronomical magnitude system mag/arcsec$^{ 2 }$ (magnitude per square arcsecond; e.g.\ \cite{berry1976,walker1970}). The idea behind this system is that if an area on the sky contained only exactly one magnitude X star in each square arcsecond, the sky brightness would be X mag/arcsec$^{ 2 }$. The magnitude system was introduced by the ancient Greek astronomer Hipparchos, who assigned a magnitude of 1 for the brightest stars visible to the naked eye, and magnitude 6 for the faintest stars visible to the naked eye (in a time before widespread light pollution). For this reason, larger values in mag/arcsec$^{ 2 }$ indicate darker skies. Once accurate measurement techniques became available, this system was later quantified and extended to stars and other objects in the sky much fainter than 6th magnitude \cite{Pogson1856}.

Astronomers measure radiances in different wavelength ranges. Similar to the way the photopic system standardized measurement of ``human visible'' light, the ``UBV system'' or ``Johnson system'' \cite{Bess1} of ultraviolet (U), blue (B), and green (visual = V) filters allows astronomers to make and report consistent observations in other color bands. The green V spectral band is not greatly different from the visual photometric spectral band, so astronomical brightness values in mag$_V$ can be approximately transformed to photometric values \cite{garstang1986}:

\begin{equation}
\label{eq:cdm2_V}
\mathrm{Luminance} \, [\mathrm{cd/m}^2] \, \approx \, 10.8 \times 10^4 \times 10^{-0.4*\mathrm{mag}_V}
\end{equation}

The darkest places on earth have a sky background brightness of about 22 mag$_V$/arcsec$^{ 2 }$, while in bright cities it is often 16--17 mag$_V$/arcsec$^{ 2 }$  \cite{kyba2015, biggs2012}. Typical sky brightness for different locations and meteorological situations are shown in Table 2. We make use of the approximation that the sky brightness is uniform across the entire sky, so the luminance and illuminance are related by a factor of $\pi$. In reality, this is a conservative estimate \cite{kocifaj2015}, as sky brightness tends to be brightest near the horizon \cite{krisciunas2010light}, especially on clear nights in predominantly artificially lit areas. The sky background brightness has also been measured from outside of the atmosphere, where the Hubble Space Telescope found 22.1--23.3 mag$_V$/arcsec$^{ 2 }$, depending on the galactic latitude (Table 3).

\begin{table}[]
\centering
\caption{Typical brightness values. These values are meant to be representative, not definitive, and the conversion between units is approximate. Luminance values are estimated for the sky at zenith based on SQM observations, except where indicated by a *. Zenith luminance and radiance values are only given where applicable (e.g.\ not during moonlit nights). Illuminances are cosine corrected.
}
\label{my-labelb}
\resizebox{\textwidth}{!}{%
\begin{tabular}{lllll}
Condition                             & surface illuminance & zenith sky luminance & zenith radiance & References                                    \\
                                                  &                     & (mcd/m$^{ 2 }$)  & (mag$_\mathrm{SQM}$/arcsec$^{ 2 }$) & \\ \hline
overcast natural night                & $<$0.6 mlux         & $<$0.2      & $>$21.8        & \cite{kyba2015, jechow2016, ribas2016}           \\
natural starlit night                 & 0.6-0.9 mlux        & 0.2-0.3     & 21.4-21.9      & See Table 3    \\
bulge of the Milky Way       & N/A                 &  2.71       &  20.5-21.0     & \cite{roach1973} (p. 23)   \\
maximum value of 50\% moon at zenith  & 25 mlux             & N/A         & N/A            & \cite{schafer1978} (p. 69)  \\
typical summer full moon              & 50-100 mlux         & N/A         & N/A            & \cite{Kyba2017} \\
maximum value of full moon at zenith  & 320 mlux            & N/A         & N/A            & \cite{Kyba2017} \\
rural night sky (clear, no moon)      &  0.7-3 mlux         &   0.25-0.8  & 20.3-21.6      & \cite{kyba2015} \\
rural night sky (overcast)            & 0.7-9 mlux          &   0.25-2.7  & 19.0-21.6      & \cite{kyba2015}    \\
suburban night sky (clear)            &   2-45 mlux         & 0.75-14     & 17.2-20.4      & \cite{kyba2015}       \\
suburban night sky (overcast)         &  6-140 mlux         &  2.1-43     & 16-19.3        & \cite{kyba2015}         \\
urban night sky (clear)               & 7-65 mlux           &   2.3-21    & 16.8-19.2      & \cite{Pun2014,kyba2015}     \\
urban night sky (overcast)            &  30-550 mlux        &  9-170      &  14.5-17.7     &  \cite{Pun2014,kyba2015}    \\
end of nautical twilight (clear)      &  8.1 mlux           &  1.9*       & 19.4           & \cite{seidelmann1992,su71115593} \\
end of civil twilight (clear)         &  3.4 lux            & 450*        &  12.9          & \cite{seidelmann1992,su71115593}   \\
DIN or CIE suggested street values    & 2-30 lux            & 0.3-2 cd/m$^{ 2 }$   & N/A   &                          \\
extremely overlit street              & 70-150 lux          & 10 cd/m$^{ 2 }$      & N/A   & \cite{kybahh2014}       \\
overcast day                          & 100-2,000 lux     & 32-640 cd/m$^{ 2 }$    & 5.6-8.8   & \cite{schlyter}      \\
direct sunlight                       & 129,000 lux         & N/A                  & N/A   & \cite{voigt2012} \\

\end{tabular}%
}
\end{table}

\begin{table}[]
\centering
\caption{{\bf Typical sky background brightness} in the astronomical Johnson V band at different astronomical observatories. If a reference is not given, the value is from \cite{Leinert1}, \cite{patat2008} or from the HST Users Handbook. The range for the Hubble Space Telescope is due to zodiacal light. {\bf It is included just as a reference of an extremely dark 'natural' sky outside the atmosphere.} Note that these are astronomical background luminances, not SQM values.}
\label{my-labelc}
\begin{tabular}{ll}
Site                                  & Zenith radiance (mag$_{V}$/arcsec$^{2}$)         \\
Tenerife, Observatory Teide           & 21.4      \\
Calar  Alto                           & 21.8         \\
La Palma, Obs. Roque de las Muchachos & 21.9        \\
Kitt Peak                             & 22.0         \\
Mauna Kea, Hawaii                     & 22.05 \\
La Silla                              & 21.7        \\
Paranal                               & 21.7 (range 21.0-22.3)   \\
Hubble Space Telescope                & 22.7  (range 22.1-23.3) \\
\end{tabular}
\end{table}

\section{Constraining the night sky brightness by visual observations}

A traditional method used to qualify the night sky quality is evaluating the ``limiting magnitude'', the magnitude of the faintest star visible to the naked eye (e.g.\ \cite{garstang2000}). The technique makes use of the contrast threshold of the human visual system: with a bright sky background, only bright stars can be seen, while against a dark background fainter stars are distinguishable. Crumey \cite{crumey14} recently derived a new formula to describe the relation between limiting magnitude and sky brightness:

\begin{equation}
\label{eq:Crumey}
\mathrm{m_\mathrm{lim}} = 0.426 \mu - 2.365 - 2.5 \mathrm{log} F
\end{equation}

with m$_{lim}$ limiting visual star magnitude (in mag), $\mu$ sky background brightness in the V band in mag$_V$/arcsec$^{ 2 }$, and a field factor $F$, typically between 1.4 and 2.4, which accounts for elements such as the observer’s experience and visual acuity. Several methods can be used to classify limiting magnitude, for example observing a sequence of stars of decreasing magnitude around the celestial North Pole, or the number of visible stars in some special star fields, which is proportional to the limiting magnitude (e.g.\ \cite{1991roggemans}). In the Globe at Night \cite{globe} and ``How many stars can we still see?'' \cite{hms} projects, observers compare their view of prominent constellations to a set of star charts with integer limiting magnitudes. This lowers the possible precision of the observations to ±1.2 magnitudes \cite{Kyba3}, but makes broad participation including non-experts possible.

The ``Loss of the Night'' app for Android and iOS devices  \cite{kybaapp} was developed to allow observers with no astronomical knowledge to make observations with higher precision. The app uses the smartphone’s inertial sensors to display a live-updating star map. It directs observers to look for individual stars in the sky, and report whether they are visible (including with averted vision), not visible, or if they cannot be observed for some other reason (e.g.\ a tree in the way). As the observer reports results for many stars of differing magnitudes, precision as good as 0.05 magnitudes is possible, and the precision of individual observations can be estimated based on the self-consistency of the data. The star catalog in the Loss of the Night app only extends to about magnitude 5.2, so its useful use is restricted to urban and suburban areas. The data from both Globe at Night and the Loss of the Night app can be displayed and evaluated at the www.myskyatnight.com website.

\section{One dimensional instruments}

The following subsections give an overview of devices that measure the sky brightness using a single channel, and typically observing only at zenith. One dimensional instruments measure the sky brightness as a sum of both the sky background brightness and the stars within the viewing field. In order to facilitate data exchange, a standard data format for acquiring data from one dimensional devices was adopted in 2012 \cite{kyba2012standard}.

\subsection{The Sky Quality Meter}

The Sky Quality Meter (SQM) was originally developed by Unihedron mainly as a tool for amateur astronomers to measure the night sky brightness at their observation sites \cite{cinzano2005}. In recent years, it has been used in a large number of studies of sky brightness (Table 1). The SQM detector consists of a solid state light-to-frequency detector (TAOS TSL237S). The spectral response encompasses the photopic eye response, but is more sensitive for shorter wavelenghts, i.e.\ more blue sensitive than a truly photopic response. Luminances are reported in the unit  mag$_{SQM}$/arcsec$^{ 2 }$, but can be converted into mcd/m$^{ 2 }$ using Eq.\ (\ref{eq:cdm2_V}) and the approximate relation: 
${mag}_\mathrm{V}$ $\approx$ ${mag}_\mathrm{SQM}$.
Note, however, that differences between the three photometric systems (SQM, Johnson V and visual photometric) can occur for different colors of the night sky \cite{de2017sky}. The SQM spectral response actually depends on four components: the sensor, an infrared filter, the lens, and any weatherproof screen, and is discussed in detail in \cite{de2017sky}.

\begin{figure}[h]
\caption{Photograph showing the Sky Quality Meter installed in its protective housing (SQM-LU left), along with an expanded view (SQM-LE right). Image and caption from \cite{Kyba2}}
\centering
\includegraphics[width=0.5\textwidth]{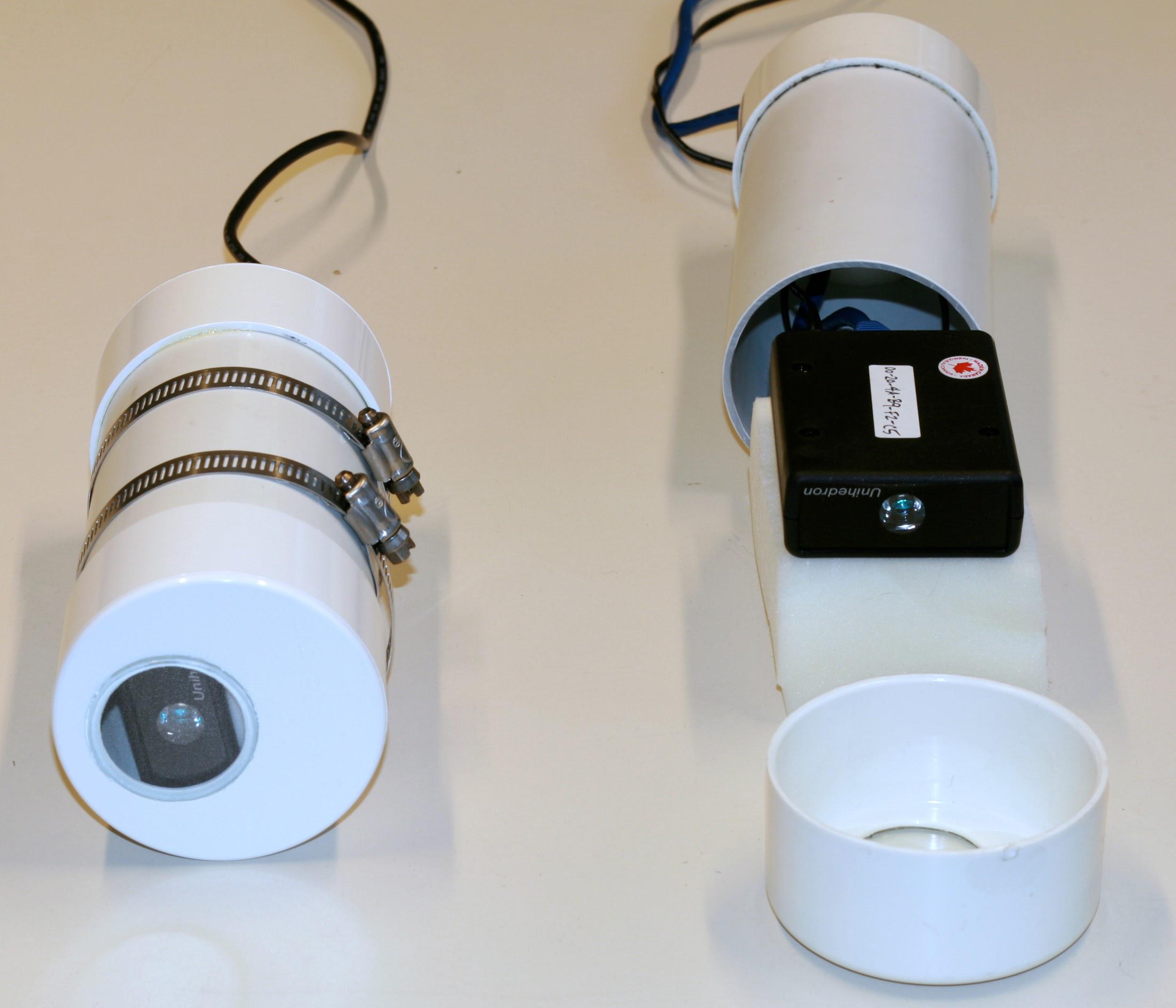}
\end{figure}

Luminance values down to the order of 100 $\mu$cd/m$^{ 2 }$ ($\sim$22.5 mag$_{SQM}$/arcsec$^{2}$) can be measured. The field of view of the SQM is 20$^{\circ}$ (full width at half maximum, FWHM) in the case of more commonly used lensed version (SQM-L). Normally measurements are taken near zenith. In areas with little artificial skyglow, nearby lamps (e.g.\ at 10$^{\circ}$ above the horizontal) can affect the measurement \cite{nolle2015}. This can be due to the residual sensitivity of the SQM at large angles, or scattered light from the lamp itself dominating over the natural sky brightness.

\begin{figure}[h]
\caption{Spectral responses for the SQM (blue, top left), the Johnson V band (blue, top right),  green band digital single lens reflex DSLR camera (blue, bottom left, see Sec 5.2), the lightmeter (blue bottom right, measured by A. Müller, see Sec 4.3). For comparison the photopic response curve (red hashed) is given.}
\centering
\includegraphics[width=1\textwidth]{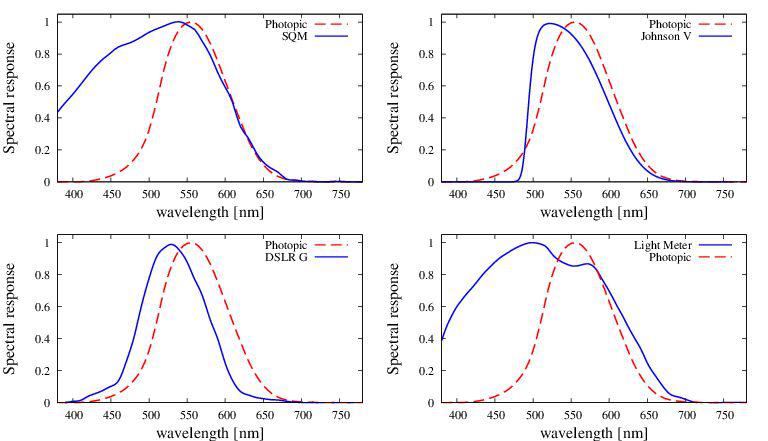}
\end{figure}

The SQMs are calibrated against a reference light meter. This reference light meter is a calibrated lux meter that meets CIE (International Commission on Illumination) regulations with an accuracy of 5\%. The applied light source was an integrating sphere using a compact fluorescent bulb until February 2011, and afterwards a green Light Emitting Diode (LED) light source with its peak at 520 nm (Tekatch, priv.\ comm.). The LED calibrated instruments (since serial number 5944) measure about 0.15 to 0.2 mag/arcsec$^{ 2 }$ systematically brighter than the older ones. The SQMs are calibrated to the same constant surface brightness of 8.71 mag/arcsec$^{ 2 }$. The dark current is determined in complete darkness. The temperature dependence of the photodiode was determined for a sample of sensors, and is corrected for by the electronics before readout \cite{schnitt2013}.

The manufacturer reports an accuracy of $\pm$10\%, corresponding to $\pm$0.1 mag$_\mathrm{SQM}$/arcsec$^{ 2 }$. During the KIck-Off IntercomparisonS Campaign (KIOS 2011), nine SQM’s operated in the Dutch Night Sky Brightness Monitoring network were intercompared at the Cabauw Experimental Site for Atmospheric Research in The Netherlands. The KIOS 2011 campaign showed an initial scatter between the individual instruments of ±14\%, ranging from −16\% to +20\%. With intercalibration methods it was possible to reduce this to 0.5\%, and −7\% to +9\%, respectively \cite{s111009603}.

The long-term behaviour of SQMs is still an issue to be examined in detail. Pioneering work in this context has been done by So \cite{so2014}, who measured a long-term change in the SQM measurements mainly related to degradation of housing window. Also den Outer et al.\ \cite{denouter2015} discuss long-term changes in the response of this kind of devices.

Different types of SQM devices are available. The most important difference between the devices is those with lens (SQM-L) and without. We strongly recommend the use of the L version because it reduces the field of view to around 20 degrees (FWHM), leading to more consistent readings when nearby sources of light are present. The SQM is available as a handheld or connected device, via USB, Ethernet or RS232. These connected devices require a computer in the field for data acquisition and storage. A data logging device (SQM-LU-DL) can store data without a computer. All devices for continuous outdoor measurements require weatherproof housing (available off-the shelf) to operate under all weather conditions. 

Since it is a rather inexpensive device, ready to use and with the possibility to do automatic measurements, the SQM is widely used for monitoring night sky brightness. In long-term observations, the sampling rate should be ideally at least once per minute \cite{kyba2012standard}, but a 15 minute interval is sometimes necessary for battery-operated SQM-LU-DL. In the case of scanning observations or relatively rapid changes in the night sky brightness, it is critical that the sampling period is longer than the time needed by the device to record an observation. User-written acquisition software enables the user to extend this range. The minimum sampling rate depends on the amount of light.

In the case of making observations with a handheld SQM-L, it is recommended to skip the first 3--4 measurements, and then to average the result of four observations, rotating the SQM (and observer’s body) 90$^{\circ}$ after each observation to a different compass direction \cite{Lonne2015}. If the SQM-L is affected by stray light, this may minimize or reveal the effect. It also reduces errors due to pointing inaccuracy \cite{Lonne2017}. If the four observations are not self-consistent (maximum range about 0.2 mag$_{SQM}$/arcsec$^{ 2 }$), then it is probably not a good location, and the data should not be recorded. It is not recommended to use SQM devices to report measurements on clear moonlit nights because of the possibility of stray light on the detector  \cite{Lonne2015,de2017sky}. Handheld SQM-L observations can be archived with Globe at Night or via the Loss of the Night app, and can be viewed at www.myskyatnight.com.

SQM-L meters have also successfully been used to sample multiple points in the celestial hemisphere \cite{birrieladkins2010}. This permits quantitative monitoring toward multiple azimuths, and thus tracking of the relative contributions from multiple light domes. Another method is the use on a car's roof together with a GPS detector, to collect positions and sky brightness simultaneously over large areas with a software called "Roadrunner" \cite{infantes2011}. 

\textbf{Recently, the European funded project STARS4ALL has developed a new detector that uses the same TSL237 photodiode detector as the SQM. This instrument is called TESS and it has an extended bandpass compared to SQM devices. TESS also uses a dichroic filter, allowing for better coverage of the red band of the visible spectra with good response \cite{zamorano2017}.}

\textbf{TESS was designed with the goal of creating a large European network with inexpensive but well tested photometers. In fact, the device is calibrated by the manufacturer, and it contains a complete system to transmit the data to STARS4ALL servers, where it will be accessible for researchers and general public. The device also includes an infrared detector, in order to obtain information about the presence or absence of clouds in the field of view of the detector.}

\subsection{Dark Sky Meter}

The Dark Sky Meter is an iPhone app which uses the back camera of iPhone 4S (and later models) to collect light and determine a sky brightness value. The app was developed by DDQ, a software company from the Netherlands, and is available via the Apple Store. Older and other models (e.g.\ iPads and iPods) are not supported because the camera chip does not detect enough light in a single exposure. The iPhone camera is not designed for long exposures, but the most recent versions of the iPhone’s CMOS (Complementary metal–oxide–semiconductor) sensor (Sony IMX145) are sensitive enough to collect light at 21 mag/arsec$^{ 2 }$, with a practical limit around 21.8 mag/arcsec$^{ 2 }$. Apple restricts developers to adjust exposure times, so readings are taken as a series of shots.

In addition to recording sky radiance, the app records device inclination (to know if the device is pointed correctly towards the zenith), moon phase, cloudiness (input by user), and GPS location. For scientific use, broadband spectral radiances are transmitted separately as three RGB values. The units of measurement are mag$_\mathrm{SQM}$/arcsec$^{ 2 }$. Estimated naked eye limiting magnitude is also reported. The sensitivity range is between 12 and 22 mag/arcsec$^{ 2 }$, with the darkest value depending on the iPhone model, as the software restricts readings above 22 mag/arcsec$^{ 2 }$. The advertised measurement accuracy is ±0.2 mag/arcsec$^{ 2 }$. The typical absolute calibration differences are 20\% for iPhone4S devices and 30\% with iPhone5 devices. The field of view is around 20$^\circ$ based on binned 240x240 LRGB images.

The data obtained can be automatically submitted to the public Globe at Night database, and are available for further works and analysis. Data under clear night conditions and with the phone pointed towards zenith are available at www.myskyatnight.com, and the complete dataset can be viewed at www.darkskymeter.com.

\subsection{Solar-cell-based Lightmeter}

The IYA-lightmeter (International Year of Astronomy 2009) uses quasi-continuous measurements of the photoelectric current of a solar cell (see Fig.\ 3) to generate proxies for illuminance and irradiance (total radiation) at a site. It is designed for long-term monitoring and uses SI-units to support communication with the public, technical lighting and the Earth- and life sciences. Because artificial night-brightening is strongest with "bad" weather and near the horizon, the instrument is all-sky and all-weather capable. Long term stability of the device is assured by the industrial solar cell, developed for outdoor use. Remaining weathering effects of the cell or the readout electronics are taken into account by repeated on-site, on-the-fly calibrations to the Sun, the Moon and the twilight and an atmospheric model \cite{muller2011,ochi2014}. These are made from the monitoring time-series and require no extra measurements or visits to the site.

To expand the dynamic range, at bright light levels the lightmeter response is non-linear. Below 0.1 lx, the lightmeter has a linear response, and can be calibrated to another instrument by simultaneously measuring a weak reference source, e.g.\ the half Moon, and applying a constant factor to the readout. The natural light of moonless, astronomical nights is dominated by the airglow and modulated by the atmosphere. The variation exceeds a factor of 2 and needs separation from artificial brightening. The high-frequency lightmeter time-series (temporal resolution: up to several samples per second) allow the use of cloud-indicators developed for astronomical- and pyranometer-time series, and, when the Sun or the Moon is present, the derivation of atmospheric extinction. The daylight capability allows to separate contributions due to atmospheric trends by comparison with existing IMO-standardised, global, long-term, total-radiation series of climate and meteorological research.

The Lightmeter has a sensitivity at a readout rate of up to 10 Hz from about 10 $\mu$lx (10$^{-5}$ lx) to above 200 000 lx with 1\% resolution (Mark 2.3 version). Thus it covers the full range of human perception of light with one sensor. The spectral response is mainly due to the amorphous Si solar cell, with contributions by the protective glass-layer. The field of view is the upper hemisphere (2$\pi$ sr), with an essentially ideal Lambertian response to contributions away from zenith, giving a FWHM of 120$^{\circ}$. It needs a data logger with USB support (netbook, Raspberry-Pi). The data format was standardized for the German Astronomical Virtual Observatory (GAVO) ``lightweather'' database \cite{gavo}.

\begin{figure}[h]
\caption{IYA Lightmeter installation at the Teide Observatory at Izaña, Tenerife (Image by Hans Deeg).}
\centering
\includegraphics[width=0.5\textwidth]{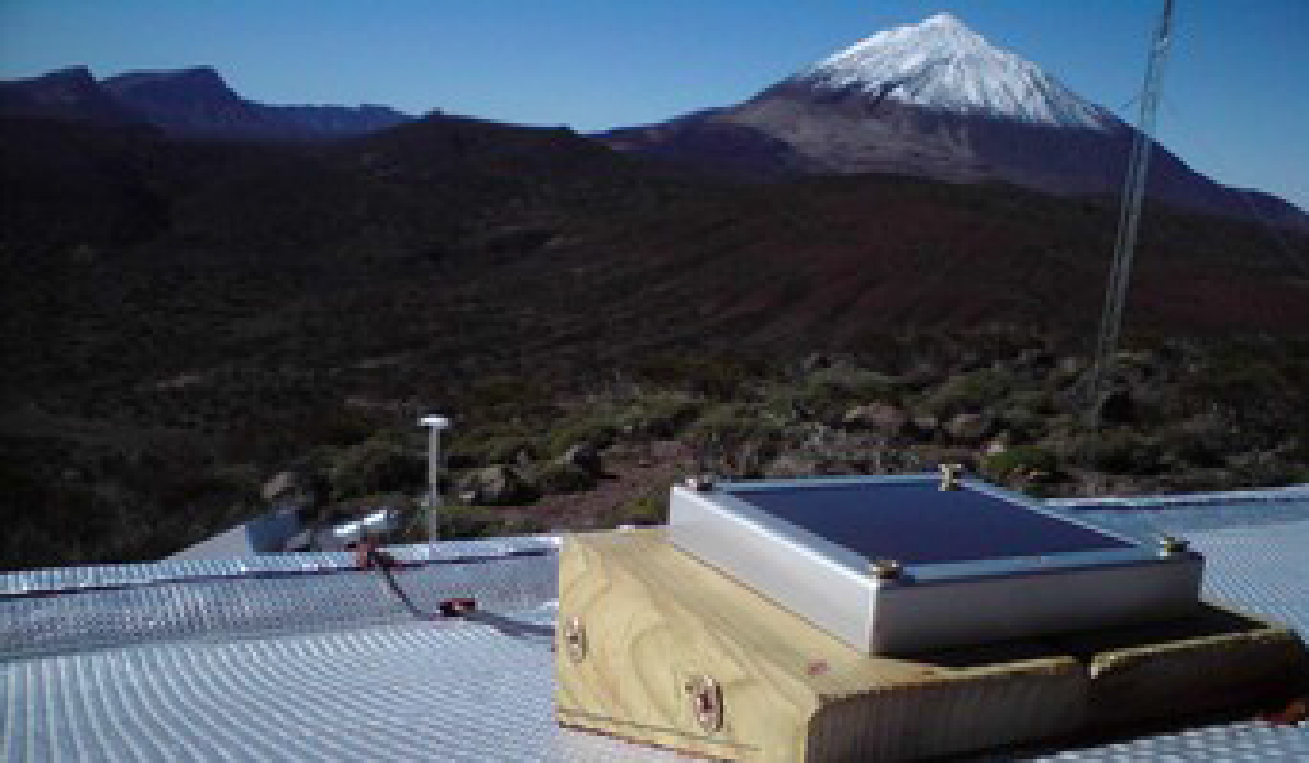}
\end{figure}

\subsection{Luxmeters and Luminance meters}

Illuminance and luminance can be measured with commercially available instruments, called luxmeters or luminance meters. Higher quality instruments are adapted to the spectral sensitivity curve of V(${\lambda}$)\cite{CIE} with an accuracy of less than 6\% and a total error of 10\% according to the German industry norm DIN 5032 class B. For an illuminance meter, the angular sensitivity characteristic must be adapted to the cosine distribution of incident light to at least 3\%. For field measurements the homogenous illuminance from the sky is difficult to realize due to obstruction by buildings, trees or disturbances through lamps. 

The main disadvantage of most such meters is the low sensitivity and low accuracy at low light levels. For a luxmeter, a limiting sensitivity of 0.1 to 0.01 lux is often specified, for luminance meters a lower limit of 0.01 to 0.001 cd/m$^{ 2 }$. Given that the night sky brightness can reach values close to 1 mlux, this implies that most luxmeters and luminance meters are not sensitive enough for the tasks discussed in this paper.

\subsection{Digilum}

``Digilum'' is a specially designed luminance meter with a large measurement range from day to night time. It was developed by Henk Spoelstra, and manufactured by Instrument Systems GmbH –- Optronik Division (Germany). The spectral sensitivity is exactly photopic (adjusted to the V$_\lambda$ sensitivity), and measures luminance between ~0.1 mcd/m$^{ 2 }$ to 20 kcd/m$^{ 2 }$. The measurement accuracy for night sky brightness values below 1 mcd/m$^{ 2 }$ ranges from 5\% up to about 20\% at 0.25 mcd/m$^{ 2 }$ for temperatures roughly above 5 $^{\circ}$C.  It is calibrated annually in the laboratory, and the dark current is corrected through a temperature correction. The main advantage of the instrument is the strict spectral sensitivity to the V(${\lambda}$) curve. It can be read out up to once per second, and the field of view is about 5$^{\circ}$. 

\section{Two dimensional (imaging) instruments }

The following subsections give an overview of devices and techniques to map and measure the sky brightness by analyzing wide-angle images (preferably images of the whole upper hemisphere). Some observations are based on CCD (charge coupled device) cameras, others on CMOS sensors of commercially available DSLR (digital single-lens reflex) cameras. Depending on the method of reduction, observations based on two dimensional instruments can report either sky luminance or sky background luminance (see Sect.\ \ref{sec:sky_brightness}). In the case of sky background luminance, they necessarily report darker values (larger mag/arcsec$^{ 2 }$) than one dimensional instruments.

\subsection{The All-Sky Transmission Monitor (ASTMON)}

\begin{figure}[h]
\caption{The ASTMON portable (Lite) version without weatherproof system (Image by Ramon Canal-Domingo).}
\centering
\includegraphics[width=0.5\textwidth]{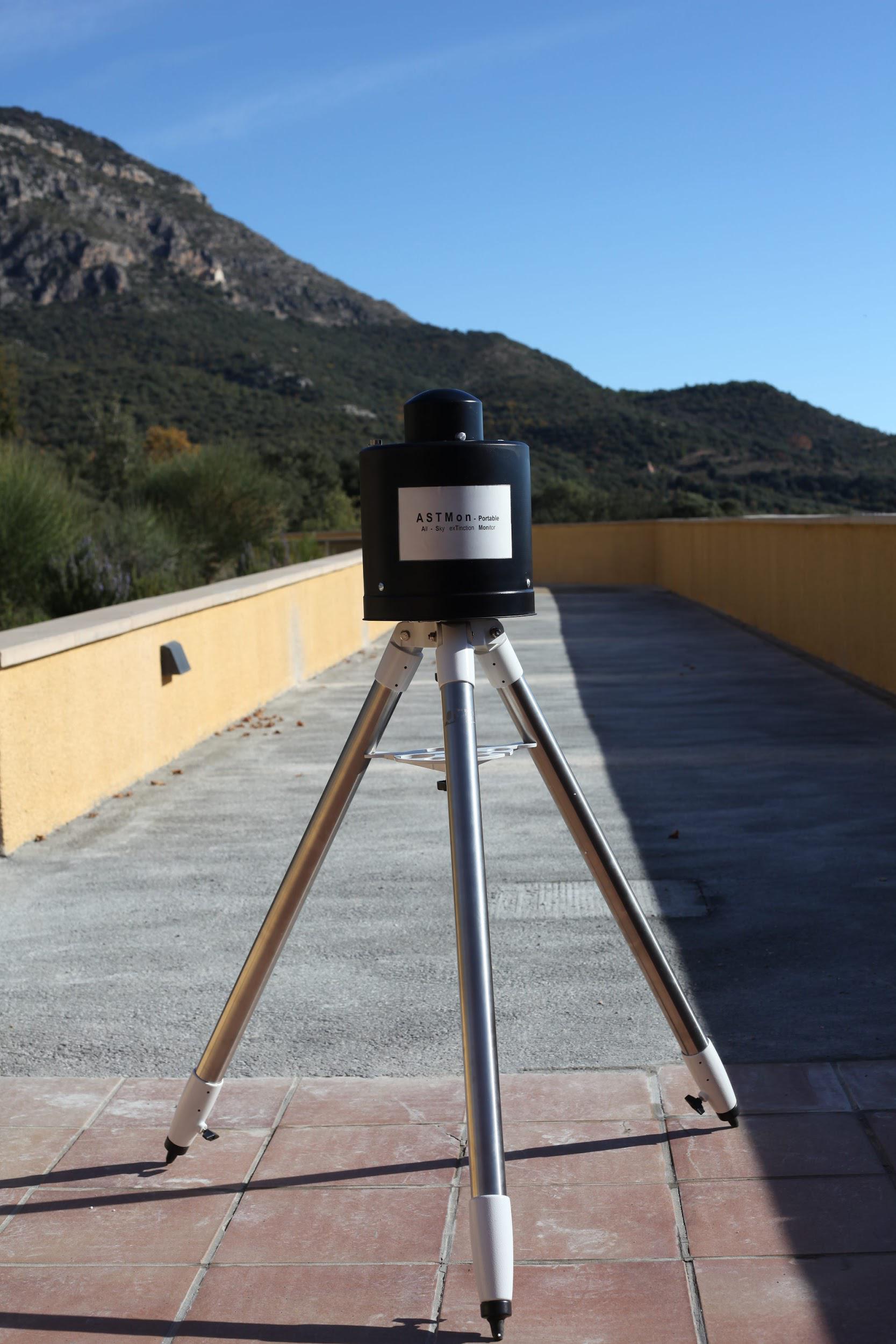}
\end{figure}

The All-Sky Transmission Monitor (ASTMON) is based on a f=4.5mm fisheye lens and an integrated astronomical CCD camera. It measures the luminance of the complete night sky in several wavelength bands including the standard astronomical unit mag$_{V}$/arcsec$^{2}$.  The system is designed to perform a continuous monitoring of the surface brightness of the night sky background in a fully robotic mode. In addition to the sky background brightness, ASTMON can provide atmospheric extinction and cloud coverage estimates for the entire sky surface at the same time \cite{aceituno2011}.

The spectral range of the instrument is directly related to the spectral sensitivity of the detector and the use of the filter wheel, which may contain one, three, or five filters depending on the version. The most common setup is with Johnson B, V, and R astronomical filters, but other filters with a 1.25 inch diameter may be added.

Because the detector is an astronomical CCD camera, different exposure times can be used to cover a wide range of sky luminances. When setting the exposure, it is recommended to keep star counts below 40,000 to avoid loss of linearity. The calibration procedure is based on astronomical photometry and the accuracy of the measurement of sky brightness is related to this procedure. Typical accuracies are around 0.15 magnitudes for U and I filters and around 0.02 for B, V and R Johnson filters, which corresponds to about 15\% and 2\%, respectively.

The instrument is available in three different versions: ASTMON Full, ASTMON Lite (see Fig.\ 4) and ASTMON Micro. The Full version is designed to be installed outdoors as a fully robotic and continuous monitoring station. This version is protected with a complete enclosure and with a solar shutter to prevent sunlight from damaging the system.  The Lite version is a portable, with a tripod and weatherproof enclosure. The system can be used outdoors for a few weeks without problems, but it is not as safe as the full permanent version. Finally the last version is Micro, which is the smallest and operates without a filter wheel, so only a Johnson V filter is permanently installed. The system is completely controlled by a computer through a specific software designed by the manufacturer of the instrument. In case of the portable devices, it is necessary to correctly adjust the device's orientation.

Every observation image has to provide a good signal-to-noise ratio and enough bright stars to find an astrometric and photometric solution. An algorithm evaluates the image to produce a catalogue of the stars in the field of view, and this catalogue is cross-matched with an astrometrically and photometrically calibrated catalogue. The processing of data includes a typical astronomical image preprocessing (dark, bias, and flat fields). This last step is especially critical because of the field distortion in an all-sky device, so ASTMON is provided with a master flat, and it is possible to update the calibration with a uniformly illuminated sphere like DomeLight \cite{aceituno2011}.

The processing of the data to obtain night sky brightness measurements is based on classical astronomical photometry, with the determination of zero points and extinction coefficient (see Appendix). In the case of bad photometric conditions without identified stars, ASTMON uses a default calibration. In case of good photometric conditions, the parameters are determined and are updated as a new default. With all the calibrations and parameters established, the system can determine the luminance of the sky background in mag/arcsec$^{ 2 }$  \cite{falchi2011} (see Fig.\ 5). The value of the sky brightness is corrected for field distortion, because in most all-sky devices, each pixel covers a different area of the sky.

Typical exposure time values in unpolluted areas are 300 seconds for the U filter and 40 seconds for the others (B, V and R). Exposures must be short enough to prevent smearing due to the rotation of the Earth, but long enough to get a good signal-to-noise ratio of the night sky background. The calibration processes are automatically run after each sequence of observations (one image with every selected filter). The typical time for a complete sequence of 5 filters is around 8 minutes, allowing the user to track nightly variations in sky brightness with 70 independent observations during a standard night. 

The software provided by the manufacturer is not open source, so some of the parameters are not controllable by the user. For this reason, there are alternative options for the processing of ASTMON images that could be applied for other all sky devices such as DSLR cameras (discussed below). One alternative option is the PyASB software \cite{nievas2003,2015IAUGA}. This open source python code is still under development, but can already generate processed all sky maps of sky background brightness, and estimates of cloud coverage and atmospheric extinction according to photometric analysis of the images. Among other places, this software has been used during the Intercomparison Campaigns of Loss of the Night Network \cite{Lonne2015,Lonne2017} and as well as in works related to all sky modal analysis decomposition  \cite{bara14}.

ASTMON has been installed in several National Parks and at astronomical observatories (e.g.\ Calar Alto Observatory, Canary Islands Observatories, Universidad Complutense de Madrid) \cite{ribas2013,ramirez2011,nievas2003}.

\begin{figure}[htbp]
\caption{Image of the sky brightness measured with ASTMON at the Montsec Astronomical Observatory by S.J.\ Ribas. The bright band across the sky is linked with galactic plane not completely subtracted by the available codes.}
\centering
\includegraphics[width=0.8\textwidth]{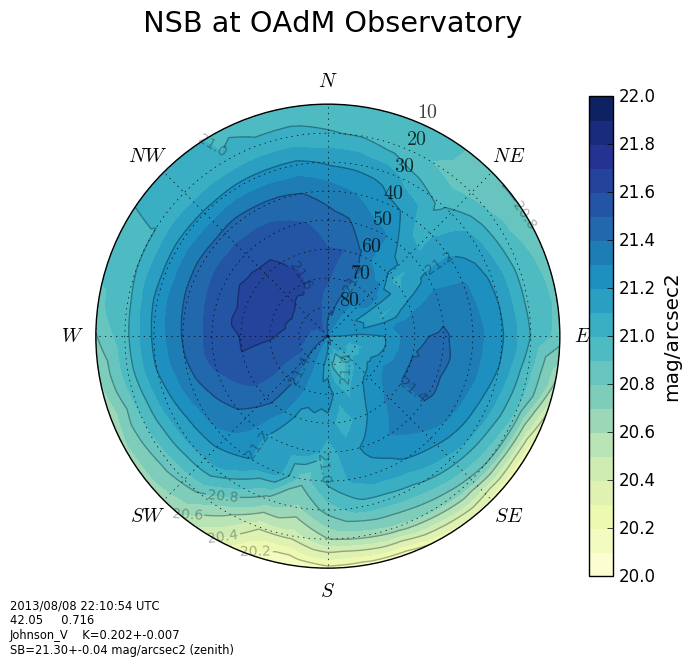}
\label{fig:astmon_allsky}
\end{figure}

\subsection{Digital cameras equipped with wide angle and fisheye lenses}

Similar to ASTMON, a commercial camera with a fisheye lens can provide quantitative information on the luminance of the whole upper hemisphere with a single exposure. Compared to CCD cameras, these systems are more easily accessible, and are highly transportable
(\cite{jechow2017a}, \cite{jechow2017b}).
The method has been proved to be an effective tool to characterize potential dark sky parks (e.g.\ \cite{kollath2010}).
Cameras cannot currently be used for measurements 'off-the-shelf', but only after calibration. The precision of the measurements depends on the calibration procedure and on the camera itself. 

The CMOS sensors and the analogue-to-digital conversion of modern cameras with 14 bit digitization provide linear measuring possibility in a 1-16384 dynamic range with a single exposure. That is usually enough for night sky monitoring, and higher dynamic range can be obtained with HDR imaging \cite{zotti2007}. For precise measurements, dark frames (images with the same exposure but covered with the lens cap) have to be taken, which can be used to compensate for the dark signal.

Calibration is the difficult part of photometry with a camera
(\cite{kocifaj+2015, lamphar2014}).
There are at least four different data processing steps involved: lens vignetting (flat fielding in CCD photometry), geometric distortion especially with wide field lenses, colour sensitivity of the camera, and the sensitivity function of the camera. 

Vignetting corrections are essential, as the transmission of a fisheye lens can drop to below 50\% at the edges compared to the image centre. The most accurate way to correct for vignetting is to use an integrating sphere in a laboratory as a ``flat'' source. With this method, a precision of $\sim$1-2\% can be obtained.

The colour sensitivity curve of the green (G) filter in cameras has a very large overlap with the astronomical Johnson V photometric band, with some degree of similarity to both the photopic and scotopic curves of human vision (see Fig.\ 2). Therefore, the most straightforward method to derive luminance from camera images is to extract the brightness values only from the G colour band in the raw data. More accurate luminance values can be derived using a linear combination of the RGB values with suitable coefficients (much smaller coefficients for the R and B compared to the G band).

In order to estimate the possible error of the colour transformation, Kolláth \cite{kollath2010} selected a set of spectra representing common light sources and used average camera sensitivity curves for the R, G, B filters and simulated different scenarios in fitting the coefficients. This showed that the relative error due to colour transformation is not larger than 3\%, and even smaller than 2\% for the Johnson V filter.

There are two different methods to calibrate the DSLR system: (a) laboratory measurements with a luminance meter and standard light sources; (b) stellar photometry to get extinction corrected sky background in astronomical V magnitude (see, e.g., \cite{kloppenborg2013,mumpuni2009,hiscocks2013}). The calibration should be performed at different luminance and exposure levels in order to map the whole dynamic range of the transmission curve. In both cases, it is difficult to reach a precision better than 5\%.

Summing the error sources of the calibration steps together, we can conclude that the precision of camera photometry can reach the 10\% range with thorough measurements and data processing. When the precision is not critical but the directional distribution of sky brightness is important (e.g.\ complementary data to other instruments), the images can be scaled to the other measurements as an alternative way of calibration \cite{haenel2015}. 

For most cameras, the longest exposure time allowed without a remote switch is 30 seconds. In bright locations, a shorter exposure time will be necessary. Test exposures are needed to ensure that the sky exposure is far from both the zero level and saturation. According to our experience, up to 30 second exposures can be used in suburban locations; but under skies without light pollution, at least 2--3 minute long exposures are recommended at ISO 800--1600. 

\begin{figure}[htbp]
\caption{\textbf{Image of the sky brightness measured with a DSLR camera a) RGB image and b) false colour plot showing the luminance. The image was taken by A.\ Jechow in the Spanish village of Ager during the 2016 Stars4all-LoNNe intercomparison campaign \cite{Lonne2017}.}}
\centering
\includegraphics[width=0.66\textwidth]{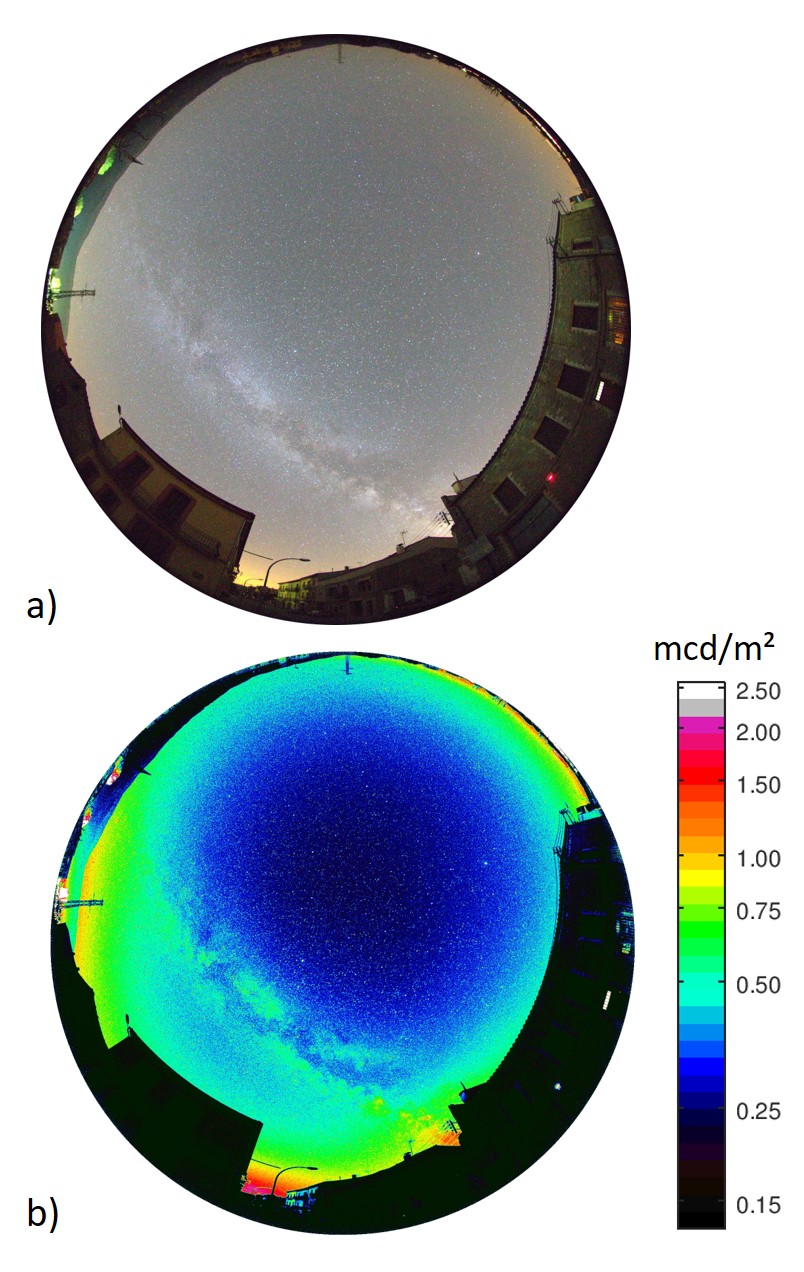}
\label{fig:DSLR}
\end{figure}

The data obtained by cameras are stored in an image format. Compressed formats like jpg and gif cannot be used, as they do not store the original values observed for each pixel. Rather, data must be stored in a raw and/or astronomical FITS format. It is common to generate false colour images of the measurements, as this provides the simplest visualization of the data. See Fig.\ \ref{fig:DSLR} for an example of an RGB image a) and a luminance false colour plot b).

In addition to the night sky brightness, the correlated colour temperature of the night sky can also be derived by analysing raw DSLR images. For this purpose and many more tasks, a dedicated software, ``Sky Quality Camera'', has been developed by Andrej Mohar (priv.\ comm.).

\subsection{All-sky mosaics}

This measurement approach pioneered by the US National Parks Service (NPS) provides all-sky coverage similar to that obtained with a fisheye lens, but by mosaicking multiple wide-field images taken in succession \cite{1538-3873-119-852-192,2016duriscoe}.  The NPS image scale (93.5 arcsec / pixel) is sufficient to perform accurate stellar photometry required for calibration to known standards, and produces panoramic or fisheye 39.3 mega-pixel images. Detailed images of sky luminance, expressed as astronomical V magnitudes per square arcsecond, can be transformed into photopic measures (e.g.\ cd/m$^{ 2 }$) or horizontal or vertical illuminance (expressed as lux) at the ground level.

\begin{figure}[htbp]
\caption{The camera system used by the US National Park Service for all-sky mosaics.}
\centering
\includegraphics[width=0.5\textwidth]{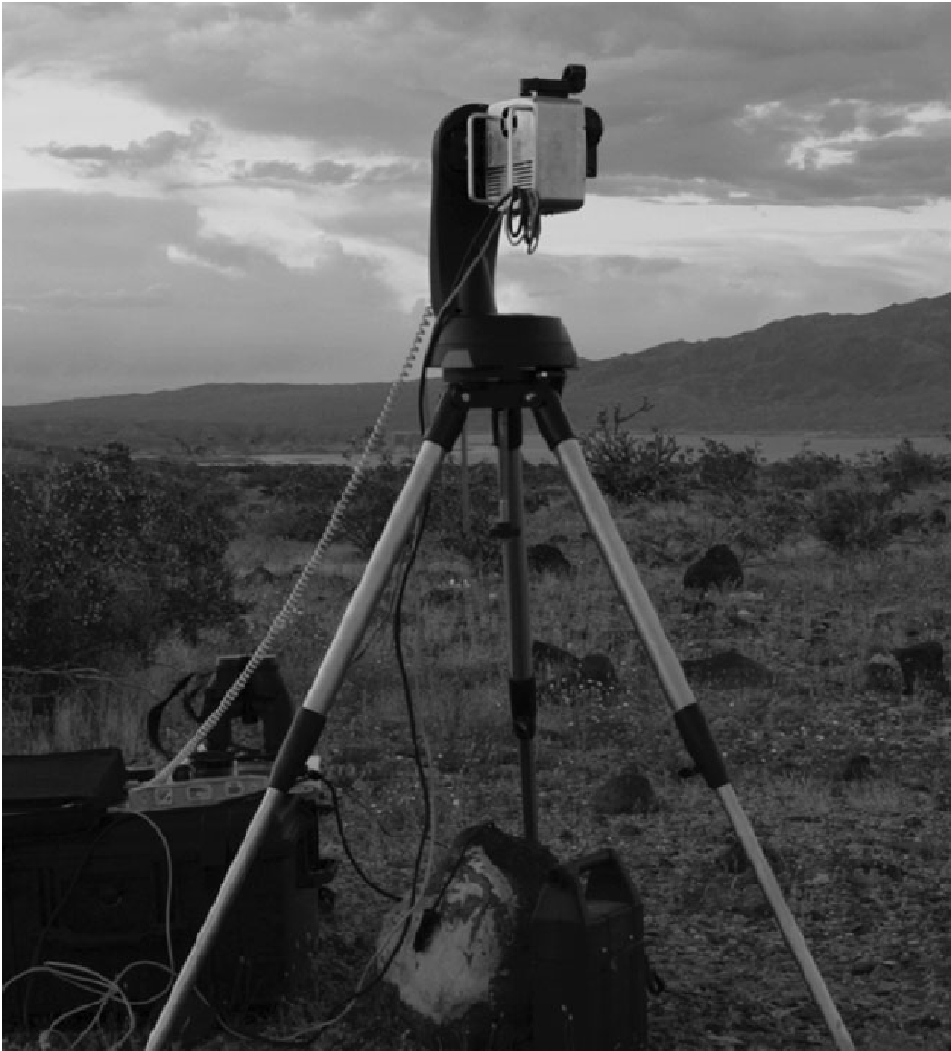}
\label{fig:US_NPS_cam}
\end{figure}

The NPS hardware is a 1024 x 1024 px CCD detector (front-illuminated Kodak 1001E), a standard 50 mm f/2 camera lens, an astronomical photometric Bessel V filter with IR blocker, and a consumer-grade computer-controlled robotic telescope mount (Fig.\ \ref{fig:US_NPS_cam}). A total of 45 images covers the entire upper hemisphere and the lower hemisphere down to -6$^\circ$, with a full acquisition requiring about 20 minutes. Data collection is orchestrated with a small portable computer, commercial software, and custom scripts. 

\begin{figure}[htbp]
\caption{False-colour depiction of the night sky brightness as derived with the US NPS camera system shown in Fig.\ \ref{fig:US_NPS_cam}}
\centering
\includegraphics[width=1\textwidth]{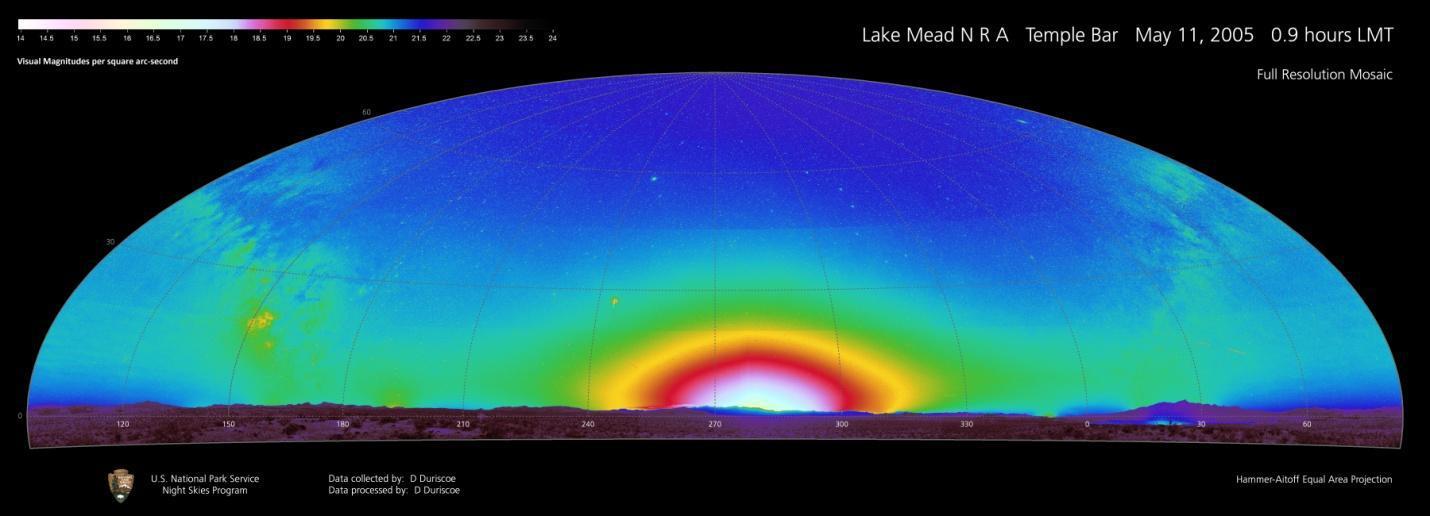}
\label{fig:US_NPS_image}
\end{figure}

The processing of the images follows the standard astronomical photometry methods, with preprocessing with master flats, bias and dark frames. Determination of zero point and extinction coefficient is done by aperture photometry with standard stars. The typical maximum error in luminance measurement across the entire frame is ±10 $\mu$cd/m$^{ 2 }$.
Wide angle rectilinear lenses produce images (see Fig.\ \ref{fig:US_NPS_image}) with a varying pixel scale from center to edge. The pixel scale used to determine the number of square arc seconds per pixel for the entire frame is that which occurs on the image where correction by the master flat is unity.
The high resolution across the celestial hemisphere obtained by a robotic system is an advantage in quantifying small distant sources of sky glow. It is also beneficial for measuring the luminance of small bright sources accurately, and for detecting change over time,  and for quantifying atmospheric extinction. The system is less suited for intensive surveillance monitoring, but well-suited for long-range monitoring from mountain tops and high-value sites.

A similar system has been employed by Falchi \cite{falchi2011} (see also \cite{2003MmSAI..74..458C} and \cite{falchi1998}). While their detector has a relatively low resolution, the large pixel size (24 $\mu$m) allows for relatively short integration times per frame to achieve adequate signal to noise in the sky background with an f/2.8 optical system, even in dark environments (18 seconds maximum exposure).  By fitting a strong neutral density filter in front of the lens, the same camera may be used to measure luminance or illuminance of nearby bright sources (such as unshielded outdoor lamps).

\section{Spectra of the night sky}

\begin{figure}[htbp]
\caption{Portable version of SAND-4 spectrometer.}
\centering
\includegraphics[width=0.5\textwidth]{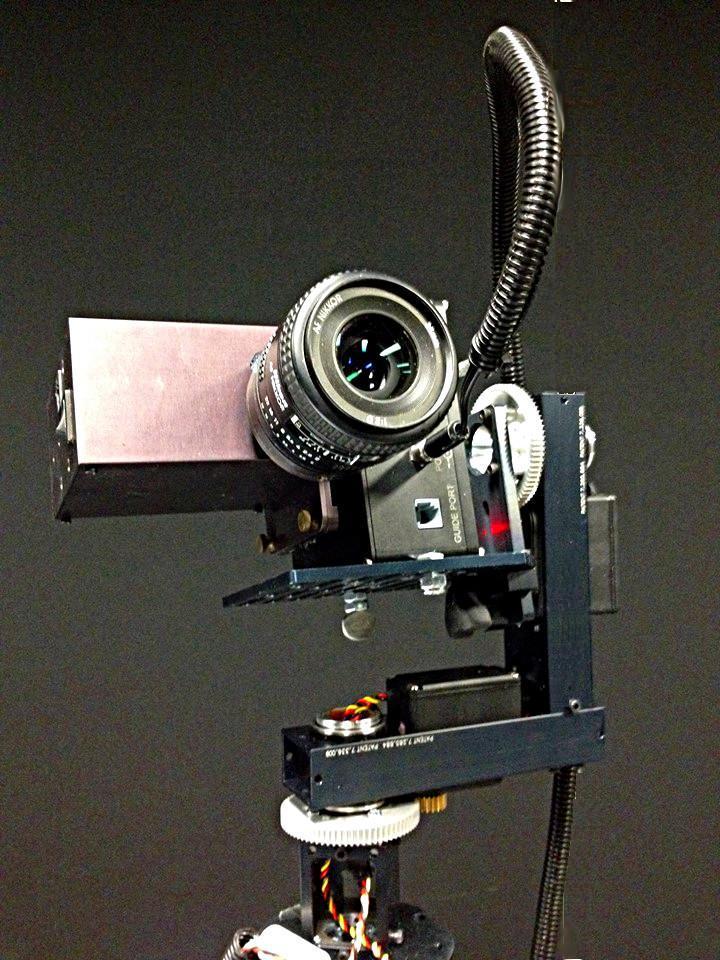}
\label{fig:SAND}
\end{figure}

This section discusses the study of the spectral power distribution (SPD) of night sky brightness. The SPD is produced by a combination of natural and artificial sources, and is dependent on nearby environmental characteristics such as atmospheric aerosol content or effects of the reflecting surfaces (ground, trees, buildings). Since artificial light sources often have considerably different SPDs, evaluation of the SPD during nighttime can provide information regarding the kind of sources that are responsible for the generation of night sky brightness. This is particularly beneficial during the current rapid change in lighting technology \cite{de2017sky}.

One approach to evaluate the SPD is the classical usage of telescopes and astronomical spectrometers to decompose the light and identify different contributions on the night sky brightness. This strategy was used, e.g., at the ESO Observatories in Chile \cite{patat2008}, in La Palma \cite{benn2007}, and at the Vienna Observatory \cite{puschnig2014night}. Another option are dedicated devices to evaluate the SPD of the night sky. The Spectrometer for Aerosol Night Detection (SAND) will be presented as an example.

SAND was initially developed for work in urban areas, but it was adapted with better sensitivity to operate in dark places \cite{Aube2007}. The latest version of SAND is shown in Fig.\ \ref{fig:SAND}. SAND-4 is a long slit spectrometer which uses a CCD imaging camera as light sensor. SAND-4 has a spectral resolution of 2 nm, and the spectral range is from 400 to 720 nm. The instrument is fully automated so that it can operate on its own with minimal human intervention.

\begin{figure}[htbp]
\caption{Spectrum of the night sky toward azimuth = 113$^{\circ}$ and zenith angle of z = 75$^{\circ}$ at El Leoncito Observatory without moon. Abbreviations: Hg (Mercury), Na (Sodium) OI (Neutral Oxygen), HPS (High Pressure Sodium)}
\centering
\includegraphics[width=1\textwidth]{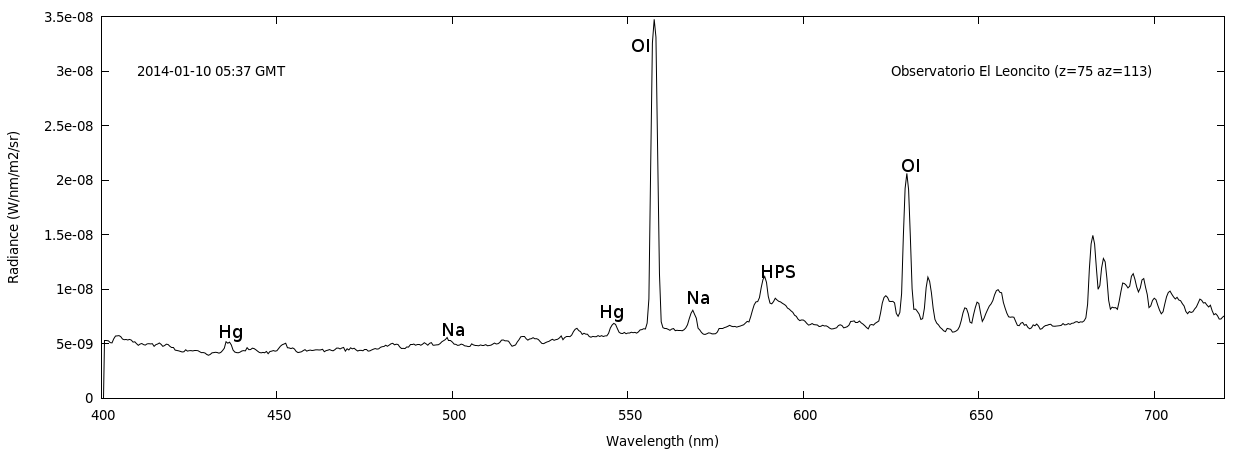}
\label{fig:SAND_ELO}
\end{figure}

\begin{figure}[htbp]
\caption{Spectrum of the night sky toward zenith (z=0) on roof of Universidad Complutense de Madrid}
\centering
\includegraphics[width=1\textwidth]{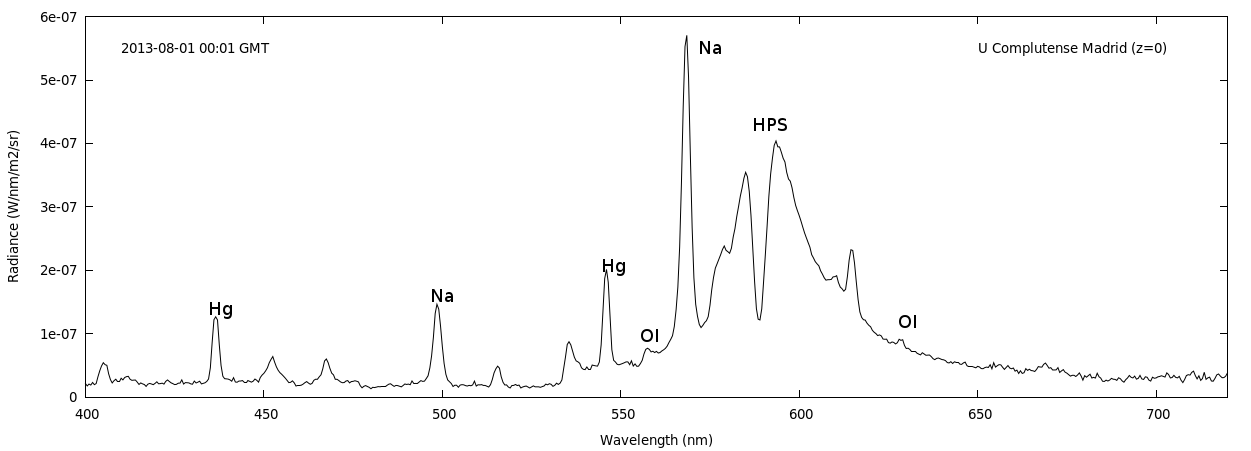}
\label{fig:SAND_UCM}
\end{figure}

With spectrometers, it is relatively easy to separate total sky radiance into its major contributing sources. Figure \ref{fig:SAND_ELO} shows a typical spectrum from an area minimally affected by artificial skyglow and without moonlight. This spectrum was taken with SAND in January 2014 at El Leoncito Observatory (Argentina) during a site characterization campaign in the framework of the Cherenkov Telescope Array project \cite{tcta2010design}. An example of a spectrum for a highly light polluted site is shown in Fig.\ \ref{fig:SAND_UCM}. In that case, artificial sky brightness is clearly dominant, and the natural contribution can be effectively neglected (at least when the moon is down). The typical integration time for urban sites is on the order of a few minutes, while the integration time rises to about two hours for sites without artificial skyglow. 

\section{Conclusions and outlook}

Quantifying light levels in a given nocturnal environment is important for many different disciplines in science, ranging from ecology to astronomy. One major obstacle for progress in this interdisciplinary field has been the large number of measurement systems and units in use. Conversion between measured quantities is roughly possible in some cases (e.g.\ mag/arcsec$^{ 2 }$ to cd/m$^{ 2 }$), while other times a conversion is only possible under particular assumptions (e.g.\ luminance cd/m$^{ 2 }$ to illuminance lux, see \cite{kocifaj2015}). The proper choice of the measurement method also depends on the information one wants to get. Is it sufficient to know the sky brightness only at the zenith, shall the illuminance of the whole sky on a horizontal surface be measured, or is it necessary to know the vertical illuminance of light sources near the horizon?

To characterize a specific location at a specific time, the ideal observation would report spectrally resolved radiance at relatively fine angular resolution in all possible directions. Unfortunately, this cannot be achieved with currently available instruments at nocturnal luminance levels typical in non-urban environments. Trade-offs must be made according to the experimental task that is being undertaken. Examples of such tasks are monitoring the long-term trend in a location \cite{so2014}, and characterizing an experimental field site \cite{jechow2016}.

During the past two decades, several new experimental techniques for measuring the night sky brightness have been developed. We have presented a number of commonly used instruments which are suitable for this purpose. These instruments can be sorted into several classes: “one-dimensional” versus “imaging” instruments, and broadband versus spectrally resolving instruments. Each of the instruments has differing strengths and weaknesses, which may make them more or less appropriate for a given experimental task. An overview of each of the instruments discussed is given in Tab.\ 4.

One dimensional instruments are the most strongly suited for monitoring temporal (especially long-term) changes in sky brightness. As many are relatively low-cost, it means that they can be used in many different locations, including regular use by citizen scientists (often amateur astronomers). Such instruments have several major weaknesses, however. First, they only measure the sky brightness but not the atmospheric extinction, which influences the astronomical sky quality considerably. Second, in most installations, they only provide information about the sky brightness in a given pointing direction (usually zenith), while real skies are characterized by strong angular gradients (Fig.\ \ref{fig:US_NPS_image}). Third, they have broadband spectral response, and this does not correspond to standards such as photometrical V or astronomical Johnson V \cite{de2017sky}. Finally, under very clear skies, the distinction between brightening of the sky by natural airglow and brightening by a moderate amount of scattered artificial light is almost impossible on the basis of measurements with such devices.

A decisive strength of imaging instruments is that they can provide information about the radiation from all directions. The data is intuitive to understand, and if directed properly the instruments can be used to provide information about what a nocturnally active animal actually perceives. Moreover, they provide broadband spectral information in three (RGB) channels. This is helpful to monitor changes in artificial lighting techniques (especially during the transition to solid state lighting \cite{de2017sky}), and spectral regions beyond the V band could be important to understand their influence on animals or plants affected by artificial light at night. Nocturnal insects, for example, are strongly influenced by ultraviolet and blue light \cite{van2014spectral}. In this respect, DSLR measurements in the B channel can deliver valuable information on the biological impact of light sources.

All-sky information can be easily obtained in a single image with 180$^\circ$ fisheye lenses. However, such lenses have strong vignetting, transmitting only about 50 percent of the light at the horizon where most artificial light sources are. To measure these influences accurately, calibration and orientation of fisheye cameras is very critical. Longer focal lengths have fewer problems with vignetting, but distortions must be well understood in order to stitch pictures together to a full sky frame. One of the barriers to making use of cameras is the lack of free software for calibration and display of the data. There has been substantial progress in this area, for example the development of software like PyASB  \cite{nievas2003}, dclum  \cite{kollath2010}, and the National Park Service display system. Nevertheless, there is a clear need for universally accepted calibration algorithms for deriving night sky luminances from camera images, like the ones used in luminance measurements (e.g.\ LMK mobile air \cite{techno}). 

For most instruments, the long term stability is not well known. Digital cameras seem to be very stable, while it is known that the plastic window of the SQM housing loses transmission over time, at least under tropical climates (So 2014). Calibration could either be done under photometric conditions in a laboratory, or using stars as a stable light source whose brightness generally is known to 1\% accuracy. 

In some cases, the decision of which instrument to use will not be either/or. For example, long-term measurements with an SQM would be greatly complemented by periodic validation with an imaging system and/or a spectrometer. It is important to periodically re-calibrate long term systems under laboratory conditions, or as part of intercomparison campaigns  \cite{Lonne2015,kyba2015,phdthesis}.

To ease interpretation of current results in the long-term future (when the currently available devices are no longer commercially available), it would be useful to establish a database of device and lens properties (e.g. geometrical correction), so that people in the future could recover the radiometric values for comparison to contemporary devices. This database would also be useful at the current time to ease calibration of cameras. It would be ideal to store some sort of ``minimally processed'' data rather than raw camera images.

\begin{table}[]
\centering
\caption{Overview of the instruments discussed in the paper.}
\label{my-labeld}
\resizebox{\textwidth}{!}{%
\begin{tabular}{l | l | l | l}
\hline
Instrument & Type of Measurement & Measurement Unit & Price \\ \hline
Visual observations & 1D photometric & limiting magnitude           & Free \\
SQM                 & 1D photometric & mag$_{SQM}$/arcsec$^{2}$     & From 150€ \\ 
Dark Sky Meter      & 1D photometric & $\sim$mag$_{V}$/arcsec$^{2}$ & Free (requires iPhone) \\ 
Luxmeter            & 1D photometric & lux (illuminance)            & From 120€ \\ 
Digilum             & 1D photometric & cd/m$^2$                     & $\sim$10,000€ \\ 
Lightmeter          & 1D photometric & W/m$^2$, $\sim$lux           & Currently sold out, IYA 2009 was 100€ \\ 
ASTMON              & 2D photometric & mag$_V$/arcsec$^2$           & From 9,000€ \\ 
DSLR+fisheye        & 2D photometric & $\sim$cd/m$^2$, $\sim$mag$_V$/arcsec$^2$ & From 1,300€ \\
NPS All sky mosaics & 2D photometric & cd/m$^2$, mag$_V$/arcsec$^2$ & $\sim$12,000€ \\ 
SAND                & spectroscopic  & W/(m$^2$nm\,sr)              & 4,000€ \\ 
\end{tabular}%
}
\end{table}

Based on our experience with each of the techniques our discussions during the development of this review, we conclude that in many situations calibrated digital cameras with fisheye lenses usually provide the best compromise between cost, ease-of-use, and amount of information obtained. The development of standard software for calibration and display of such data should be a high priority of the field of light at night researchers.

\section*{Appendix. Classical Astronomical Photometry}

The classical approach to evaluate the night sky background brightness is based on astronomical star photometry in defined photometric systems (eg. UBV Johnson system). Normally this technique is used for stellar brightness measurements with photometers and CCD detectors through filters on telescopes. Some new devices use the same strategy to determine the sky brightness, taking standard astronomical stars as reference candles to estimate the background brightness. Here we will summarize the basic steps of this procedure.

First of all, in order to use this method it is necessary to observe during nighttime (after the astronomical twilight when the Sun is below 18$^\circ$ under the horizon). To obtain good photometric results, it is also necessary to observe when the Moon is below 10° under the horizon \cite{falchi2011}.

The best quality results are obtained during nights of photometric quality. These nights follow the conditions of stability that results in a constant ratio between atmospheric extinction ($\kappa$) and airmass ($\chi$). The airmass can be summarized as the total amount of air crossed by the light from a star as it propagates through the atmosphere, and it is directly determined by the altitude above the horizon of the selected star as follows (Bouguer method):

\begin{equation}
\label{airmass}
\chi = \mathrm{sec} \  z \  (1 - 0.0012 \ \mathrm{tan}^{2} \  z)
\end{equation}

where $z$ is the zenith angle (90$^\circ$ - altitude) of the selected star in the moment of the observation.

The observation strategy starts with the observation of standard astronomical stars placed at different airmasses, to cover the whole range of altitude above horizon for which we expect to determine the sky brightness. For best results, it is necessary to use standard stars of different astronomical magnitude. For each field of observation we will determine the airmass for the standard stars in each measurement, and the digital number counts detected with our astronomical device. These counts could be determined simply using the classical double circle technique \cite{falchi2011} to measure the number of counts of the star without sky contribution, and divide this number of counts by the exposure time to get real counts per second values. Otherwise, every other method of stellar photometry can be used, and a large range of image processing and specialized software is available for different operation systems (Sextractor, Aperture Photometry, DAOPHOT, Iris, AstroImageJ, Astroart, Midas, IRAF).

The standard stars have the value of their brightness (magnitude) recorded in photometric catalogues. Combining the catalogue magnitude data ($mag$) with the airmass ($\chi$) and the number of counts ($I$) per second that we detect, it is possible to determine the instrumental zero points (ZP) and the extinction coefficient ($\kappa$) using the classical equation \cite{falchi2011,aceituno2011}:

\begin{equation}
\label{zp_equation}
2.5 \, \mathrm{log} _{10} \ I \ = \mathrm{ZP} - \kappa \cdot \chi
\end{equation}

The determination of ZP and $\kappa$ is established using the set of all of the standard astronomical stars observed during the night. If the atmosphere is stable and there are no changes in the instruments, these two parameters remain constant during the observation period.

Finally, using the background counts per second, we can evaluate the sky brightness of any image obtained during the night, taking into account the image scale. The best option is to determine the number of counts inside a defined rectangle that does not contain visible stars. The number of counts has to be divided by the angular area of the rectangle and by exposure time to get ``counts per second and per arcsec$^{2}$'' ($I_{sky}$) \cite{falchi2011}. 
The sky brightness (SB) is finally determined in magnitudes per arcsec$^{2}$ for any selected position of the sky using the determined parameters with the equation:

\begin{equation}
\label{sb_equation}
\mathrm{SB} \ = \mathrm{ZP} - \ 2.5 \, \mathrm{log} _{10} \ I _{sky}
\end{equation}

This method has been intensively used by Wim Schmidt to measure the sky background brightness in extended regions of the Netherlands \cite{schmidt2011}.

\section*{Acknowledgements}

This article is based upon work from COST Action ES1204 LoNNe (Loss of the Night Network), supported by COST (European Cooperation in Science and Technology). Initial drafts were written during COST sponsored short term scientific missions of Zoltán Kolláth and Salvador J. Ribas to Austria and Germany. We gratefully acknowledge the help of Franz Binder (Vienna) with the formatting of this paper. We thank Anne Aulsebrook and Travis Longcore for comments on a draft version.


\section*{References}
\begin{thebibliography}{92}
\providecommand{\natexlab}[1]{#1}
\providecommand{\url}[1]{\texttt{#1}}
\providecommand{\href}[2]{#2}
\providecommand{\path}[1]{#1}
\providecommand{\eprint}[1]{\href{http://arxiv.org/abs/#1}{\path{#1}}}
\providecommand{\DOIprefix}{doi:}
\providecommand{\ArXivprefix}{arXiv:}
\providecommand{\URLprefix}{URL: }
\providecommand{\Pubmedprefix}{pmid:}
\providecommand{\doi}[1]{\href{http://dx.doi.org/#1}{\path{#1}}}
\providecommand{\Pubmed}[1]{\href{pmid:#1}{\path{#1}}}
\providecommand{\BIBand}{and}
\providecommand{\bibinfo}[2]{#2}
\ifx\xfnm\undefined \def\xfnm[#1]{\unskip,\space#1}\fi
\bibitem[{Longcore and Rich(2004)}]{long1}
\bibinfo{author}{Longcore\xfnm[ T.]}, \bibinfo{author}{Rich\xfnm[ C.]}.
\newblock \bibinfo{title}{Ecological light pollution}.
\newblock \bibinfo{journal}{Frontiers in Ecology and the Environment}
  \bibinfo{year}{2004};\bibinfo{volume}{2}(\bibinfo{number}{4}):\bibinfo{pages}{191--198}.
\newblock \URLprefix
  \url{http://dx.doi.org/10.1890/1540-9295(2004)002[0191:ELP]2.0.CO;2}.
  \DOIprefix\doi{10.1890/1540-9295(2004)002[0191:ELP]2.0.CO;2}.
\bibitem[{H{\"o}lker et~al.(2016)H{\"o}lker, Wolter, Perkin and
  Tockner}]{HolkerWolter1}
\bibinfo{author}{H{\"o}lker\xfnm[ F.]}, \bibinfo{author}{Wolter\xfnm[ C.]},
  \bibinfo{author}{Perkin\xfnm[ E.K.]}, \bibinfo{author}{Tockner\xfnm[ K.]}.
\newblock \bibinfo{title}{Light pollution as a biodiversity threat}.
\newblock \bibinfo{journal}{Trends in Ecology \& Evolution}
  \bibinfo{year}{2016};\bibinfo{volume}{25}(\bibinfo{number}{12}):\bibinfo{pages}{681--682}.
\newblock \URLprefix \url{http://dx.doi.org/10.1016/j.tree.2010.09.007}.
  \DOIprefix\doi{10.1016/j.tree.2010.09.007}.
\bibitem[{Stevens et~al.(2013)Stevens, Brainard, Blask, Lockley and
  Motta}]{stevens2013adverse}
\bibinfo{author}{Stevens\xfnm[ R.G.]}, \bibinfo{author}{Brainard\xfnm[ G.C.]},
  \bibinfo{author}{Blask\xfnm[ D.E.]}, \bibinfo{author}{Lockley\xfnm[ S.W.]},
  \bibinfo{author}{Motta\xfnm[ M.E.]}.
\newblock \bibinfo{title}{Adverse health effects of nighttime lighting:
  comments on {American} medical association policy statement}.
\newblock \bibinfo{journal}{American journal of preventive medicine}
  \bibinfo{year}{2013};\bibinfo{volume}{45}(\bibinfo{number}{3}):\bibinfo{pages}{343--346}.
\bibitem[{H{\"o}lker et~al.(2010)H{\"o}lker, Moss, Griefahn, Kloas, Voigt,
  Henckel et~al.}]{holker2010}
\bibinfo{author}{H{\"o}lker\xfnm[ F.]}, \bibinfo{author}{Moss\xfnm[ T.]},
  \bibinfo{author}{Griefahn\xfnm[ B.]}, \bibinfo{author}{Kloas\xfnm[ W.]},
  \bibinfo{author}{Voigt\xfnm[ C.C.]}, \bibinfo{author}{Henckel\xfnm[ D.]},
  et~al.
\newblock \bibinfo{title}{The {Dark} {Side} of {Light}: {A} {Transdisciplinary}
  {Research} {Agenda} for {Light} {Pollution} {Policy}}.
\newblock \bibinfo{journal}{Ecology and Society}
  \bibinfo{year}{2010};\bibinfo{volume}{15}(\bibinfo{number}{4}):\bibinfo{pages}{13}.
\bibitem[{S{\'a}nchez~de Miguel et~al.(2014)S{\'a}nchez~de Miguel, Zamorano,
  G{\'o}mez~Casta{\~n}o and Pascual}]{miguel2014}
\bibinfo{author}{S{\'a}nchez~de Miguel\xfnm[ A.]},
  \bibinfo{author}{Zamorano\xfnm[ J.]},
  \bibinfo{author}{G{\'o}mez~Casta{\~n}o\xfnm[ J.]},
  \bibinfo{author}{Pascual\xfnm[ S.]}.
\newblock \bibinfo{title}{Evolution of the energy consumed by street lighting
  in {Spain} estimated with {DMSP}{-OLS} data}.
\newblock \bibinfo{journal}{J Quant Spectrosc Ra}
  \bibinfo{year}{2014};\bibinfo{volume}{139}:\bibinfo{pages}{109--117}.
\bibitem[{Miller et~al.(2013)Miller, Straka, Mills, Elvidge, Lee, Solbrig
  et~al.}]{rs5126717}
\bibinfo{author}{Miller\xfnm[ S.D.]}, \bibinfo{author}{Straka\xfnm[ W.]},
  \bibinfo{author}{Mills\xfnm[ S.P.]}, \bibinfo{author}{Elvidge\xfnm[ C.D.]},
  \bibinfo{author}{Lee\xfnm[ T.F.]}, \bibinfo{author}{Solbrig\xfnm[ J.]},
  et~al.
\newblock \bibinfo{title}{{Illuminating} the {Capabilities} of the {Suomi}
  {National} {Polar}-{Orbiting} {Partnership} {(NPP)} {Visible} {Infrared}
  {Imaging} {Radiometer} {Suite} {(VIIRS)} {Day}/{Night} {Band}}.
\newblock \bibinfo{journal}{Remote Sensing}
  \bibinfo{year}{2013};\bibinfo{volume}{5}(\bibinfo{number}{12}):\bibinfo{pages}{6717}.
\newblock \URLprefix \url{http://www.mdpi.com/2072-4292/5/12/6717}.
  \DOIprefix\doi{10.3390/rs5126717}.
\bibitem[{{S{\'a}nchez de Miguel} et~al.(2014){S{\'a}nchez de Miguel},
  {Castaño}, {Zamorano}, {Pascual}, {Ángeles}, {Cayuela}
  et~al.}]{Miguel01082014}
\bibinfo{author}{{S{\'a}nchez de Miguel}\xfnm[ A.]},
  \bibinfo{author}{{Castaño}\xfnm[ J.G.]}, \bibinfo{author}{{Zamorano}\xfnm[
  J.]}, \bibinfo{author}{{Pascual}\xfnm[ S.]},
  \bibinfo{author}{{Ángeles}\xfnm[ M.]}, \bibinfo{author}{{Cayuela}\xfnm[
  L.]}, et~al.
\newblock \bibinfo{title}{Atlas of astronaut photos of {Earth} at night}.
\newblock \bibinfo{journal}{Astron Geophy}
  \bibinfo{year}{2014};\bibinfo{volume}{55}(\bibinfo{number}{4}):\bibinfo{pages}{4.36}.
\newblock \DOIprefix\doi{10.1093/astrogeo/atu165}.
\bibitem[{De~Almeida et~al.(2014)De~Almeida, Santos, Paolo and
  Quicheron}]{de2014solid}
\bibinfo{author}{De~Almeida\xfnm[ A.]}, \bibinfo{author}{Santos\xfnm[ B.]},
  \bibinfo{author}{Paolo\xfnm[ B.]}, \bibinfo{author}{Quicheron\xfnm[ M.]}.
\newblock \bibinfo{title}{Solid state lighting review -- {Potential} and
  challenges in {Europe}}.
\newblock \bibinfo{journal}{Renewable and Sustainable Energy Reviews}
  \bibinfo{year}{2014};\bibinfo{volume}{34}:\bibinfo{pages}{30--48}.
\bibitem[{Cinzano(2005)}]{cinzano2005}
\bibinfo{author}{Cinzano\xfnm[ P.]}.
\newblock \bibinfo{title}{Night sky photometry with sky quality meter}.
\newblock \bibinfo{journal}{ISTIL Internal Rep}
  \bibinfo{year}{2005};\bibinfo{volume}{9}.
\bibitem[{{Koll{\'a}th}(2010)}]{kollath2010}
\bibinfo{author}{{Koll{\'a}th}\xfnm[ Z.]}.
\newblock \bibinfo{title}{Measuring and modelling light pollution at the
  {Zselic} {Starry} {Sky} {Park}}.
\newblock \bibinfo{journal}{Journal of Physics: Conference Series}
  \bibinfo{year}{2010};\bibinfo{volume}{218}(\bibinfo{number}{1}):\bibinfo{pages}{012001}.
\newblock \URLprefix \url{http://stacks.iop.org/1742-6596/218/i=1/a=012001}.
\bibitem[{{Cinzano} et~al.(2001{\natexlab{a}}){Cinzano}, {Falchi} and
  {Elvidge}}]{Cinzano2001}
\bibinfo{author}{{Cinzano}\xfnm[ P.]}, \bibinfo{author}{{Falchi}\xfnm[ F.]},
  \bibinfo{author}{{Elvidge}\xfnm[ C.D.]}.
\newblock \bibinfo{title}{The first {World} {Atlas} of the artificial night sky
  brightness}.
\newblock \bibinfo{journal}{Monthly Notices of the Royal Astronomical Society}
  \bibinfo{year}{2001}{\natexlab{a}};\bibinfo{volume}{328}(\bibinfo{number}{3}):\bibinfo{pages}{689--707}.
\newblock \DOIprefix\doi{10.1046/j.1365-8711.2001.04882.x}.
\bibitem[{Swaddle et~al.(2015)Swaddle, Francis, Barber, Cooper, Kyba, Dominoni
  et~al.}]{swaddle2015framework}
\bibinfo{author}{Swaddle\xfnm[ J.P.]}, \bibinfo{author}{Francis\xfnm[ C.D.]},
  \bibinfo{author}{Barber\xfnm[ J.R.]}, \bibinfo{author}{Cooper\xfnm[ C.B.]},
  \bibinfo{author}{Kyba\xfnm[ C.C.]}, \bibinfo{author}{Dominoni\xfnm[ D.M.]},
  et~al.
\newblock \bibinfo{title}{A framework to assess evolutionary responses to
  anthropogenic light and sound}.
\newblock \bibinfo{journal}{Trends Ecol Evol}
  \bibinfo{year}{2015};\bibinfo{volume}{30}(\bibinfo{number}{9}):\bibinfo{pages}{550--560}.
\bibitem[{Pendoley et~al.(2012)Pendoley, Verveer, Kahlon, Savage, Ryan
  et~al.}]{pendoley2012}
\bibinfo{author}{Pendoley\xfnm[ K.L.]}, \bibinfo{author}{Verveer\xfnm[ A.]},
  \bibinfo{author}{Kahlon\xfnm[ A.]}, \bibinfo{author}{Savage\xfnm[ J.]},
  \bibinfo{author}{Ryan\xfnm[ R.T.]}, et~al.
\newblock \bibinfo{title}{A novel technique for monitoring light pollution}.
\newblock In: \bibinfo{booktitle}{{International} {Conference} on {Health},
  {Safety} and {Environment} in {Oil} and {Gas} {Exploration} and
  {Production}}. \bibinfo{organization}{Society of Petroleum Engineers};
  \bibinfo{year}{2012},.
\bibitem[{{Jechow} et~al.(2016){Jechow}, {H{\"o}lker}, {Koll{\'a}th}, {Gessner}
  and {Kyba}}]{jechow2016}
\bibinfo{author}{{Jechow}\xfnm[ A.]}, \bibinfo{author}{{H{\"o}lker}\xfnm[ F.]},
  \bibinfo{author}{{Koll{\'a}th}\xfnm[ Z.]}, \bibinfo{author}{{Gessner}\xfnm[
  M.O.]}, \bibinfo{author}{{Kyba}\xfnm[ C.C.M.]}.
\newblock \bibinfo{title}{{Evaluating the summer night sky brightness at a
  research field site on {Lake} {Stechlin} in northeastern {Germany}}}.
\newblock \bibinfo{journal}{J Quant Spectrosc Ra}
  \bibinfo{year}{2016};\bibinfo{volume}{181}:\bibinfo{pages}{24--32}.
\newblock \DOIprefix\doi{10.1016/j.jqsrt.2016.02.005}.
  \href{http://arxiv.org/abs/1602.04537}{\tt arXiv:1602.04537}.
\bibitem[{CIE ISO23539:2005(E)/ CIE S 010/E:2004(2005)}]{CIE}
CIE ISO23539:2005(E)/ CIE S 010/E:2004.
\newblock \bibinfo{title}{{Photometry} – {The} {CIE} {System} of {Physical}
  {Photometry.}}
\newblock \bibinfo{type}{Standard}; CIE; \bibinfo{address}{Vienna, Austria};
  \bibinfo{year}{2005}.
\bibitem[{{Leinert} et~al.(1998){Leinert}, {Bowyer}, {Haikala}, {Hanner},
  {Hauser}, {Levasseur-Regourd} et~al.}]{Leinert1}
\bibinfo{author}{{Leinert}\xfnm[ C.]}, \bibinfo{author}{{Bowyer}\xfnm[ S.]},
  \bibinfo{author}{{Haikala}\xfnm[ L.K.]}, \bibinfo{author}{{Hanner}\xfnm[
  M.S.]}, \bibinfo{author}{{Hauser}\xfnm[ M.G.]},
  \bibinfo{author}{{Levasseur-Regourd}\xfnm[ A.C.]}, et~al.
\newblock \bibinfo{title}{The 1997 reference of diffuse night sky brightness}.
\newblock \bibinfo{journal}{Astron Astrophys}
  \bibinfo{year}{1998};\bibinfo{volume}{127}:\bibinfo{pages}{1--99}.
\newblock \DOIprefix\doi{10.1051/aas:1998105}.
\bibitem[{{Noll} et~al.(2012){Noll}, {Kausch}, {Barden}, {Jones}, {Szyszka},
  {Kimeswenger} et~al.}]{noll2012}
\bibinfo{author}{{Noll}\xfnm[ S.]}, \bibinfo{author}{{Kausch}\xfnm[ W.]},
  \bibinfo{author}{{Barden}\xfnm[ M.]}, \bibinfo{author}{{Jones}\xfnm[ A.M.]},
  \bibinfo{author}{{Szyszka}\xfnm[ C.]}, \bibinfo{author}{{Kimeswenger}\xfnm[
  S.]}, et~al.
\newblock \bibinfo{title}{An atmospheric radiation model for {Cerro} {Paranal}
  - {I.} {The} {optical} {spectral} {range?}}
\newblock \bibinfo{journal}{Astron Astrophys}
  \bibinfo{year}{2012};\bibinfo{volume}{543}:\bibinfo{pages}{A92}.
\newblock \URLprefix \url{http://dx.doi.org/10.1051/0004-6361/201219040}.
  \DOIprefix\doi{10.1051/0004-6361/201219040}.
\bibitem[{Chant(1935)}]{chant1935}
\bibinfo{author}{Chant\xfnm[ C.]}.
\newblock \bibinfo{title}{Sky-glow from large cities}.
\newblock \bibinfo{journal}{Journal of the Royal Astronomical Society of
  Canada} \bibinfo{year}{1935};\bibinfo{volume}{29}:\bibinfo{pages}{79}.
\bibitem[{Garstang(1986)}]{garstang1986}
\bibinfo{author}{Garstang\xfnm[ R.H.]}.
\newblock \bibinfo{title}{Model for artificial night-sky illumination}.
\newblock \bibinfo{journal}{Publications of the Astronomical Society of the
  Pacific}
  \bibinfo{year}{1986};\bibinfo{volume}{98}(\bibinfo{number}{601}):\bibinfo{pages}{364}.
\bibitem[{{Aub{\'e}}(2015)}]{Aub20140117}
\bibinfo{author}{{Aub{\'e}}\xfnm[ M.]}.
\newblock \bibinfo{title}{Physical behaviour of anthropogenic light propagation
  into the nocturnal environment}.
\newblock \bibinfo{journal}{Philosophical Transactions of the Royal Society of
  London B: Biological Sciences}
  \bibinfo{year}{2015};\bibinfo{volume}{370}(\bibinfo{number}{1667}).
\newblock \URLprefix
  \url{http://rstb.royalsocietypublishing.org/content/370/1667/20140117}.
  \DOIprefix\doi{10.1098/rstb.2014.0117}.
\bibitem[{{Cinzano} et~al.(2001{\natexlab{b}}){Cinzano}, {Falchi} and
  {Elvidge}}]{2001MNRAS}
\bibinfo{author}{{Cinzano}\xfnm[ P.]}, \bibinfo{author}{{Falchi}\xfnm[ F.]},
  \bibinfo{author}{{Elvidge}\xfnm[ C.D.]}.
\newblock \bibinfo{title}{Naked-eye star visibility and limiting magnitude
  mapped from {DMSP}-{OLS} satellite data}.
\newblock \bibinfo{journal}{Mon Not R Astron Soc}
  \bibinfo{year}{2001}{\natexlab{b}};\bibinfo{volume}{323}:\bibinfo{pages}{34--46}.
\newblock \DOIprefix\doi{10.1046/j.1365-8711.2001.04213.x}.
  \href{http://arxiv.org/abs/astro-ph/0011310}{\tt arXiv:astro-ph/0011310}.
\bibitem[{Kyba et~al.(2011{\natexlab{a}})Kyba, Ruhtz, Fischer and
  H{\"o}lker}]{kyba2011}
\bibinfo{author}{Kyba\xfnm[ C.C.]}, \bibinfo{author}{Ruhtz\xfnm[ T.]},
  \bibinfo{author}{Fischer\xfnm[ J.]}, \bibinfo{author}{H{\"o}lker\xfnm[ F.]}.
\newblock \bibinfo{title}{Lunar skylight polarization signal polluted by urban
  lighting}.
\newblock \bibinfo{journal}{Journal of Geophysical Research: Atmospheres}
  \bibinfo{year}{2011}{\natexlab{a}};\bibinfo{volume}{116}(\bibinfo{number}{D24}).
\bibitem[{Kyba et~al.(2015{\natexlab{a}})Kyba, Tong, Bennie, Birriel, Birriel,
  Cool et~al.}]{kyba2015}
\bibinfo{author}{Kyba\xfnm[ C.C.M.]}, \bibinfo{author}{Tong\xfnm[ K.P.]},
  \bibinfo{author}{Bennie\xfnm[ J.]}, \bibinfo{author}{Birriel\xfnm[ I.]},
  \bibinfo{author}{Birriel\xfnm[ J.J.]}, \bibinfo{author}{Cool\xfnm[ A.]},
  et~al.
\newblock \bibinfo{title}{Worldwide variations in artificial skyglow}.
\newblock \bibinfo{journal}{Scientific Reports}
  \bibinfo{year}{2015}{\natexlab{a}};\bibinfo{volume}{5}:\bibinfo{pages}{8409}.
\newblock \URLprefix \url{http://dx.doi.org/10.1038/srep08409}.
\bibitem[{{Ribas}(2016)}]{phdthesis}
\bibinfo{author}{{Ribas}\xfnm[ S.J.]}.
\newblock \bibinfo{title}{Caracterització de la contaminació lumínica en
  zones protegides i urbanes}.
\newblock Ph.D. thesis; Universitat de Barcelona; \bibinfo{year}{2016}.
\bibitem[{Ribas et~al.(2016)Ribas, Torra, Figueras, Paricio and
  Canal-Domingo}]{ribas2016}
\bibinfo{author}{Ribas\xfnm[ S.J.]}, \bibinfo{author}{Torra\xfnm[ J.]},
  \bibinfo{author}{Figueras\xfnm[ F.]}, \bibinfo{author}{Paricio\xfnm[ S.]},
  \bibinfo{author}{Canal-Domingo\xfnm[ R.]}.
\newblock \bibinfo{title}{How {Clouds} are {Amplifying} (or not) the {Effects}
  of {ALAN}}.
\newblock \bibinfo{journal}{International Journal of Sustainable Lighting}
  \bibinfo{year}{2016};\bibinfo{volume}{35}:\bibinfo{pages}{32--39}.
\bibitem[{{Berry}(1976)}]{berry1976}
\bibinfo{author}{{Berry}\xfnm[ R.L.]}.
\newblock \bibinfo{title}{{Light} {Pollution} in {Southern} {Ontario}}.
\newblock \bibinfo{journal}{J Roy Astron Soc Can}
  \bibinfo{year}{1976};\bibinfo{volume}{70}:\bibinfo{pages}{97--124}.
\bibitem[{{Walker}(1970)}]{walker1970}
\bibinfo{author}{{Walker}\xfnm[ M.F.]}.
\newblock \bibinfo{title}{{{The} {California} {Site} {Survey}}}.
\newblock \bibinfo{journal}{Publ Astron Soc Pac}
  \bibinfo{year}{1970};\bibinfo{volume}{82}:\bibinfo{pages}{672--698}.
\bibitem[{Pogson(1856)}]{Pogson1856}
\bibinfo{author}{Pogson\xfnm[ N.]}.
\newblock \bibinfo{title}{Magnitudes of {Thirty}-six of the {Minor} {Planets}
  for the first day of each month of the year 1857}.
\newblock \bibinfo{journal}{Mon Not R Astron Soc}
  \bibinfo{year}{1856};\bibinfo{volume}{17}:\bibinfo{pages}{12--15}.
\bibitem[{{Bessell}(2005)}]{Bess1}
\bibinfo{author}{{Bessell}\xfnm[ M.S.]}.
\newblock \bibinfo{title}{{Standard} {Photometric} {Systems}}.
\newblock \bibinfo{journal}{Annual Review of Astronomy and Astrophysics}
  \bibinfo{year}{2005};\bibinfo{volume}{43}(\bibinfo{number}{1}):\bibinfo{pages}{293--336}.
\newblock \DOIprefix\doi{10.1146/annurev.astro.41.082801.100251}.
\bibitem[{{Biggs} et~al.(2012){Biggs}, {Fouch{\'e}}, {Bilki} and
  {Zadnik}}]{biggs2012}
\bibinfo{author}{{Biggs}\xfnm[ J.D.]}, \bibinfo{author}{{Fouch{\'e}}\xfnm[
  T.]}, \bibinfo{author}{{Bilki}\xfnm[ F.]}, \bibinfo{author}{{Zadnik}\xfnm[
  M.G.]}.
\newblock \bibinfo{title}{Measuring and mapping the night sky brightness of
  {Perth}, {Western} {Australia}}.
\newblock \bibinfo{journal}{Mon Not R Astron Soc}
  \bibinfo{year}{2012};\bibinfo{volume}{421}(\bibinfo{number}{2}):\bibinfo{pages}{1450--1464}.
\bibitem[{Kocifaj et~al.(2015)Kocifaj, Posch and Solano~Lamphar}]{kocifaj2015}
\bibinfo{author}{Kocifaj\xfnm[ M.]}, \bibinfo{author}{Posch\xfnm[ T.]},
  \bibinfo{author}{Solano~Lamphar\xfnm[ H.]}.
\newblock \bibinfo{title}{On the relation between zenith sky brightness and
  horizontal illuminance}.
\newblock \bibinfo{journal}{Monthly Notices of the Royal Astronomical Society}
  \bibinfo{year}{2015};\bibinfo{volume}{446}(\bibinfo{number}{3}):\bibinfo{pages}{2895--2901}.
\bibitem[{Krisciunas et~al.(2010)Krisciunas, Bogglio, Sanhueza and
  Smith}]{krisciunas2010light}
\bibinfo{author}{Krisciunas\xfnm[ K.]}, \bibinfo{author}{Bogglio\xfnm[ H.]},
  \bibinfo{author}{Sanhueza\xfnm[ P.]}, \bibinfo{author}{Smith\xfnm[ M.G.]}.
\newblock \bibinfo{title}{Light pollution at high zenith angles, as measured at
  {Cerro} {Tololo} {Inter-American Observatory}}.
\newblock \bibinfo{journal}{Publications of the Astronomical Society of the
  Pacific}
  \bibinfo{year}{2010};\bibinfo{volume}{122}(\bibinfo{number}{889}):\bibinfo{pages}{373}.
\bibitem[{Roach and Gorden(1973)}]{roach1973}
\bibinfo{author}{Roach\xfnm[ F.E.]}, \bibinfo{author}{Gorden\xfnm[ J.L.]}.
\newblock \bibinfo{title}{The light of the night sky}.
\newblock \bibinfo{publisher}{Springer Science \& Business Media};
  \bibinfo{year}{1973}.
\bibitem[{Sch{\"a}fer(1978)}]{schafer1978}
\bibinfo{author}{Sch{\"a}fer\xfnm[ H.]}.
\newblock \bibinfo{title}{{Astronomische} {Probleme} und ihre physikalischen
  {Grundlagen}}.
\newblock \bibinfo{publisher}{Springer}; \bibinfo{year}{1978}.
\bibitem[{Kyba et~al.(2017)Kyba, Posch and Mohar}]{Kyba2017}
\bibinfo{author}{Kyba\xfnm[ C.C.M.]}, \bibinfo{author}{Posch\xfnm[ T.]},
  \bibinfo{author}{Mohar\xfnm[ A.]}.
\newblock \bibinfo{title}{How bright is full moonlight?}
\newblock \bibinfo{journal}{Astron Geophys}
  \bibinfo{year}{2017};\bibinfo{volume}{58}:\bibinfo{pages}{1.31--1.32}.
\bibitem[{{Pun} et~al.(2014){Pun}, {So}, {Leung} and {Wong}}]{Pun2014}
\bibinfo{author}{{Pun}\xfnm[ C.S.J.]}, \bibinfo{author}{{So}\xfnm[ C.W.]},
  \bibinfo{author}{{Leung}\xfnm[ W.Y.]}, \bibinfo{author}{{Wong}\xfnm[ C.F.]}.
\newblock \bibinfo{title}{{Contributions of artificial lighting sources on
  light pollution in {Hong} {Kong} measured through a night sky brightness
  monitoring network}}.
\newblock \bibinfo{journal}{J Quant Spectrosc Ra}
  \bibinfo{year}{2014};\bibinfo{volume}{139}:\bibinfo{pages}{90--108}.
\newblock \DOIprefix\doi{10.1016/j.jqsrt.2013.12.014}.
\bibitem[{{Seidelmann}(1992)}]{seidelmann1992}
\bibinfo{author}{{Seidelmann}\xfnm[ K.]}.
\newblock \bibinfo{title}{{Explanatory} {Supplement} to the {Astronomical}
  {Almanac}}.
\newblock \bibinfo{publisher}{University Science Books}; \bibinfo{year}{1992}.
\bibitem[{Holzhauer et~al.(2015)Holzhauer, Franke, Kyba, Manfrin, Klenke, Voigt
  et~al.}]{su71115593}
\bibinfo{author}{Holzhauer\xfnm[ S.I.J.]}, \bibinfo{author}{Franke\xfnm[ S.]},
  \bibinfo{author}{Kyba\xfnm[ C.C.M.]}, \bibinfo{author}{Manfrin\xfnm[ A.]},
  \bibinfo{author}{Klenke\xfnm[ R.]}, \bibinfo{author}{Voigt\xfnm[ C.C.]},
  et~al.
\newblock \bibinfo{title}{Out of the {Dark}: {Establishing} a {Large}-{Scale}
  {Field} {Experiment} to {Assess} the {Effects} of {Artificial} {Light} at
  {Night} on {Species} and {Food} {Webs}}.
\newblock \bibinfo{journal}{Sustainability}
  \bibinfo{year}{2015};\bibinfo{volume}{7}(\bibinfo{number}{11}):\bibinfo{pages}{15593}.
\bibitem[{Kyba et~al.(2014)Kyba, H{\"a}nel and Holker}]{kybahh2014}
\bibinfo{author}{Kyba\xfnm[ C.C.M.]}, \bibinfo{author}{H{\"a}nel\xfnm[ A.]},
  \bibinfo{author}{Holker\xfnm[ F.]}.
\newblock \bibinfo{title}{Redefining efficiency for outdoor lighting}.
\newblock \bibinfo{journal}{Energy Environ Sci}
  \bibinfo{year}{2014};\bibinfo{volume}{7}:\bibinfo{pages}{1806--1809}.
\newblock \URLprefix \url{http://dx.doi.org/10.1039/C4EE00566J}.
  \DOIprefix\doi{10.1039/C4EE00566J}.
\bibitem[{Schlyter(2009)}]{schlyter}
\bibinfo{author}{Schlyter\xfnm[ P.]}.
\newblock \bibinfo{title}{Radiometry and photometry in astronomy.}
\newblock \bibinfo{howpublished}{\url{http://stjarnhimlen.se/comp/radfaq}};
  \bibinfo{year}{2009}.
\newblock \bibinfo{note}{Accessed: 3 Jan 2017}.
\bibitem[{{Voigt}(2012)}]{voigt2012}
\bibinfo{author}{{Voigt}\xfnm[ H.]}.
\newblock \bibinfo{title}{Abriss der {Astronomie}}.
\newblock \bibinfo{publisher}{John Wiley \& Sons}; \bibinfo{year}{2012}.
\bibitem[{{Patat}(2008)}]{patat2008}
\bibinfo{author}{{Patat}\xfnm[ F.]}.
\newblock \bibinfo{title}{The dancing sky: 6 years of night-sky observations at
  {Cerro} {Paranal}}.
\newblock \bibinfo{journal}{Astron Astrophys}
  \bibinfo{year}{2008};\bibinfo{volume}{481}(\bibinfo{number}{2}):\bibinfo{pages}{575--591}.
\newblock \URLprefix \url{http://dx.doi.org/10.1051/0004-6361:20079279}.
  \DOIprefix\doi{10.1051/0004-6361:20079279}.
\bibitem[{{Garstang}(2000)}]{garstang2000}
\bibinfo{author}{{Garstang}\xfnm[ R.H.]}.
\newblock \bibinfo{title}{Limiting visual magnitude and night sky brightness}.
\newblock \bibinfo{journal}{Mem Soc Astron It}
  \bibinfo{year}{2000};\bibinfo{volume}{71}:\bibinfo{pages}{83--92}.
\bibitem[{{Crumey}(2014)}]{crumey14}
\bibinfo{author}{{Crumey}\xfnm[ A.]}.
\newblock \bibinfo{title}{Human contrast threshold and astronomical
  visibility}.
\newblock \bibinfo{journal}{Mon Not R Astron Soc}
  \bibinfo{year}{2014};\bibinfo{volume}{442}:\bibinfo{pages}{2600--2619}.
\bibitem[{Roggemans(1991)}]{1991roggemans}
\bibinfo{author}{Roggemans\xfnm[ P.]}.
\newblock \bibinfo{title}{{Watching} the {Perseid} {Meteor} {Shower}}.
\newblock \bibinfo{journal}{Sky \& Telescope}
  \bibinfo{year}{1991};\bibinfo{volume}{August}:\bibinfo{pages}{172--173}.
\bibitem[{glo(2016)}]{globe}
\bibinfo{title}{Globe at night.}
\newblock \bibinfo{howpublished}{\url{http://www.globeatnight.org}};
  \bibinfo{year}{2016}.
\newblock \bibinfo{note}{Accessed:2 Jan 2016}.
\bibitem[{hms(2001)}]{hms}
\bibinfo{title}{{Kuffner-Sternwarte}, {How} {many} {stars...?}}
\newblock \bibinfo{howpublished}{\url{http://starbright.astronomy2009.at/}
  Archived by WebCite\textregistered at http://www.webcitation.org/6oT9ZcjPf};
  \bibinfo{year}{2001}.
\newblock \bibinfo{note}{Accessed: 22 Feb 2017}.
\bibitem[{Kyba et~al.(2013)Kyba, Wagner, Kuechly, Walker, Elvidge, Falchi
  et~al.}]{Kyba3}
\bibinfo{author}{Kyba\xfnm[ C.C.M.]}, \bibinfo{author}{Wagner\xfnm[ J.M.]},
  \bibinfo{author}{Kuechly\xfnm[ H.U.]}, \bibinfo{author}{Walker\xfnm[ C.E.]},
  \bibinfo{author}{Elvidge\xfnm[ C.D.]}, \bibinfo{author}{Falchi\xfnm[ F.]},
  et~al.
\newblock \bibinfo{title}{{Citizen} {Science} {Provides} {Valuable} {Data} for
  {Monitoring} {Global} {Night} {Sky} {Luminance}}.
\newblock \bibinfo{journal}{Scientific Reports}
  \bibinfo{year}{2013};\bibinfo{volume}{3}:\bibinfo{pages}{1835 EP --}.
\newblock \URLprefix \url{http://dx.doi.org/10.1038/srep01835}.
\bibitem[{Kyba(2015)}]{kybaapp}
\bibinfo{author}{Kyba\xfnm[ C.C.]}.
\newblock \bibinfo{title}{Brief introduction to the loss of the night app
  project}.
\newblock
  \bibinfo{howpublished}{\url{http://lossofthenight.blogspot.co.at/2015/01/brief-introduction-to-loss-of-night-app.html}};
  \bibinfo{year}{2015}.
\newblock \bibinfo{note}{Accessed:9 Feb 2016}.
\bibitem[{Kyba and Lolkema(2012)}]{kyba2012standard}
\bibinfo{author}{Kyba\xfnm[ C.C.M.]}, \bibinfo{author}{Lolkema\xfnm[ D.E.]}.
\newblock \bibinfo{title}{A community standard for recording skyglow data}.
\newblock \bibinfo{journal}{Astronomy \& Geophysics}
  \bibinfo{year}{2012};\bibinfo{volume}{53}(\bibinfo{number}{6}):\bibinfo{pages}{6--17}.
\bibitem[{{S{\'a}nchez de Miguel} et~al.(2017){S{\'a}nchez de Miguel},
  {Aub{\'e}}, {Zamorano}, {Kocifaj}, {Roby} and {Tapia}}]{de2017sky}
\bibinfo{author}{{S{\'a}nchez de Miguel}\xfnm[ A.]},
  \bibinfo{author}{{Aub{\'e}}\xfnm[ M.]}, \bibinfo{author}{{Zamorano}\xfnm[
  J.]}, \bibinfo{author}{{Kocifaj}\xfnm[ M.]}, \bibinfo{author}{{Roby}\xfnm[
  J.]}, \bibinfo{author}{{Tapia}\xfnm[ C.]}.
\newblock \bibinfo{title}{{Sky} {Quality} {Meter} measurements in a colour
  changing world}.
\newblock \bibinfo{journal}{Mon Not R Astron Soc}
  \bibinfo{year}{2017};\bibinfo{volume}{467}:\bibinfo{pages}{2966--2979}.
\bibitem[{Kyba et~al.(2011{\natexlab{b}})Kyba, Ruhtz, Fischer and
  Hölker}]{Kyba2}
\bibinfo{author}{Kyba\xfnm[ C.C.M.]}, \bibinfo{author}{Ruhtz\xfnm[ T.]},
  \bibinfo{author}{Fischer\xfnm[ J.]}, \bibinfo{author}{Hölker\xfnm[ F.]}.
\newblock \bibinfo{title}{{Cloud} {Coverage} {Acts} as an {Amplifier} for
  {Ecological} {Light} {Pollution} in {Urban} {Ecosystems}}.
\newblock \bibinfo{journal}{PLOS ONE}
  \bibinfo{year}{2011}{\natexlab{b}};\bibinfo{volume}{6}(\bibinfo{number}{3}):\bibinfo{pages}{1--9}.
\newblock \DOIprefix\doi{10.1371/journal.pone.0017307}.
\bibitem[{Nolle(2015)}]{nolle2015}
\bibinfo{author}{Nolle\xfnm[ M.]}.
\newblock \bibinfo{title}{{Einfluss} von hellen {Lichtquellen} auf {Messungen}
  mit dem {SQM-L}}.
\newblock \bibinfo{journal}{Journal für Astronomie, Vereinigung der
  Sternfreunde}
  \bibinfo{year}{2015};\bibinfo{volume}{54}(\bibinfo{number}{3}):\bibinfo{pages}{21--23}.
\bibitem[{Schnitt et~al.(2013)Schnitt, Ruhtz, Fischer, H{\"o}lker and
  Kyba}]{schnitt2013}
\bibinfo{author}{Schnitt\xfnm[ S.]}, \bibinfo{author}{Ruhtz\xfnm[ T.]},
  \bibinfo{author}{Fischer\xfnm[ J.]}, \bibinfo{author}{H{\"o}lker\xfnm[ F.]},
  \bibinfo{author}{Kyba\xfnm[ C.C.M.]}.
\newblock \bibinfo{title}{Temperature {Stability} of the {Sky} {Quality}
  {Meter}}.
\newblock \bibinfo{journal}{Sensors}
  \bibinfo{year}{2013};\bibinfo{volume}{13}(\bibinfo{number}{9}):\bibinfo{pages}{12166}.
\newblock \URLprefix \url{http://www.mdpi.com/1424-8220/13/9/12166}.
  \DOIprefix\doi{10.3390/s130912166}.
\bibitem[{{den Outer} et~al.(2011){den Outer}, {Lolkema}, {Haaima}, {Hoff},
  {Spoelstra} and {Schmidt}}]{s111009603}
\bibinfo{author}{{den Outer}\xfnm[ P.]}, \bibinfo{author}{{Lolkema}\xfnm[ D.]},
  \bibinfo{author}{{Haaima}\xfnm[ M.]}, \bibinfo{author}{{Hoff}\xfnm[ R.v.d.]},
  \bibinfo{author}{{Spoelstra}\xfnm[ H.]}, \bibinfo{author}{{Schmidt}\xfnm[
  W.]}.
\newblock \bibinfo{title}{{Intercomparisons} of {Nine} {Sky} {Brightness}
  {Detectors}}.
\newblock \bibinfo{journal}{Sensors}
  \bibinfo{year}{2011};\bibinfo{volume}{11}(\bibinfo{number}{10}):\bibinfo{pages}{9603}.
\newblock \URLprefix \url{http://www.mdpi.com/1424-8220/11/10/9603}.
  \DOIprefix\doi{10.3390/s111009603}.
\bibitem[{{So}(2014)}]{so2014}
\bibinfo{author}{{So}\xfnm[ C.W.]}.
\newblock \bibinfo{title}{Observational studies of contributions of artificial
  and natural light factors to the night sky brightness measured through a
  monitoring network in {Hong} {Kong}}.
\newblock Ph.D. thesis; University of Hong Kong; \bibinfo{year}{2014}.
\bibitem[{den Outer et~al.(2015)den Outer, Lolkema, Haaima, van~der Hoff,
  Spoelstra and Schmidt}]{denouter2015}
\bibinfo{author}{den Outer\xfnm[ P.]}, \bibinfo{author}{Lolkema\xfnm[ D.]},
  \bibinfo{author}{Haaima\xfnm[ M.]}, \bibinfo{author}{van~der Hoff\xfnm[ R.]},
  \bibinfo{author}{Spoelstra\xfnm[ H.]}, \bibinfo{author}{Schmidt\xfnm[ W.]}.
\newblock \bibinfo{title}{Stability of the {Nine} {Sky} {Quality} {Meters} in
  the {Dutch} night sky brightness monitoring network}.
\newblock \bibinfo{journal}{Sensors}
  \bibinfo{year}{2015};\bibinfo{volume}{15}(\bibinfo{number}{4}):\bibinfo{pages}{9466--9480}.
\bibitem[{Kyba et~al.(2015{\natexlab{b}})Kyba, Bouroussis, Canal-Domingo,
  Falchi, Giacomelli, H{\"a}nel et~al.}]{Lonne2015}
\bibinfo{author}{Kyba\xfnm[ C.C.M.]}, \bibinfo{author}{Bouroussis\xfnm[ C.]},
  \bibinfo{author}{Canal-Domingo\xfnm[ R.]}, \bibinfo{author}{Falchi\xfnm[
  F.]}, \bibinfo{author}{Giacomelli\xfnm[ A.]},
  \bibinfo{author}{H{\"a}nel\xfnm[ A.]}, et~al.
\newblock \bibinfo{title}{Report of the 2015 {LoNNe} {Intercomparison}
  {Campaign}} \bibinfo{year}{2015}{\natexlab{b}};\URLprefix
  \url{http://gfzpublic.gfz-potsdam.de/pubman/faces/viewItemOverviewPage.jsp?itemId=escidoc:1124000}.
\bibitem[{Ribas et~al.(2017)Ribas, Aub{\'e}, Bar{\'a}, Bouroussis,
  Canal-Domingo, Espey et~al.}]{Lonne2017}
\bibinfo{author}{Ribas\xfnm[ S.J.]}, \bibinfo{author}{Aub{\'e}\xfnm[ M.]},
  \bibinfo{author}{Bar{\'a}\xfnm[ S.]}, \bibinfo{author}{Bouroussis\xfnm[ C.]},
  \bibinfo{author}{Canal-Domingo\xfnm[ R.]}, \bibinfo{author}{Espey\xfnm[ B.]},
  et~al.
\newblock \bibinfo{title}{Report of the 2016 {STARS4ALL}/{LoNNe}
  {Intercomparison} {Campaign}} \bibinfo{year}{2017};\URLprefix
  \url{http://doi.org/10.2312/GFZ.1.4.2017.001}.
\bibitem[{Birriel and Adkins(2010)}]{birrieladkins2010}
\bibinfo{author}{Birriel\xfnm[ J.]}, \bibinfo{author}{Adkins\xfnm[ J.K.]}.
\newblock \bibinfo{title}{{A} {Simple}, {Portable} {Apparatus} to {Measure}
  {Night} {Sky} {Brightness} at {Various} {Zenith} {Angles}}.
\newblock \bibinfo{journal}{Journal of the American Association of Variable
  Star Observers}
  \bibinfo{year}{2010};\bibinfo{volume}{38}:\bibinfo{pages}{221--229}.
\bibitem[{Rosa~Infantes(2011)}]{infantes2011}
\bibinfo{author}{Rosa~Infantes\xfnm[ D.]}.
\newblock \bibinfo{title}{The {Road} {Runner} {System}}.
\newblock \bibinfo{journal}{4th International Symposium for Dark Sky Parks,
  Montsec, 2011} \bibinfo{year}{2011};\URLprefix
  \url{http://darkskyparks.splet.arnes.si/files/2011/09/RoadRunner.pdf};
  \bibinfo{note}{accessed: 14 Apr 2017}.
\bibitem[{Zamorano et~al.(2017)Zamorano, Garc{\'\i}a, Tapia, de~Miguel, Pascual
  and Gallego}]{zamorano2017}
\bibinfo{author}{Zamorano\xfnm[ J.]}, \bibinfo{author}{Garc{\'\i}a\xfnm[ C.]},
  \bibinfo{author}{Tapia\xfnm[ C.]}, \bibinfo{author}{de~Miguel\xfnm[ A.S.]},
  \bibinfo{author}{Pascual\xfnm[ S.]}, \bibinfo{author}{Gallego\xfnm[ J.]}.
\newblock \bibinfo{title}{Stars4all night sky brightness photometer}.
\newblock \bibinfo{journal}{International Journal of Sustainable Lighting}
  \bibinfo{year}{2017};\bibinfo{volume}{18}:\bibinfo{pages}{49--54}.
\bibitem[{M{\"u}ller et~al.(2011)M{\"u}ller, Wuchterl and Sarazin}]{muller2011}
\bibinfo{author}{M{\"u}ller\xfnm[ A.]}, \bibinfo{author}{Wuchterl\xfnm[ G.]},
  \bibinfo{author}{Sarazin\xfnm[ M.]}.
\newblock \bibinfo{title}{Measuring the night sky brightness with the
  lightmeter}.
\newblock \bibinfo{journal}{Revista Mexicana de Astronoma y Astrofisica}
  \bibinfo{year}{2011};\bibinfo{volume}{41}:\bibinfo{pages}{46--49}.
\bibitem[{{Ochi} and {Wuchterl}(2014)}]{ochi2014}
\bibinfo{author}{{Ochi}\xfnm[ N.]}, \bibinfo{author}{{Wuchterl}\xfnm[ G.]}.
\newblock \bibinfo{title}{Long-term measurement of the night sky brightness in
  {Japan} using {Lightmeters}: 2009-2012 data}.
\newblock \bibinfo{journal}{Journal of Tokyo University, Natural Science}
  \bibinfo{year}{2014};(\bibinfo{number}{58}):\bibinfo{pages}{1--12}.
\bibitem[{gav(2009)}]{gavo}
\bibinfo{title}{{GAVO} ({German} {Astrophysical} {Virtual} {Observatory})
  {Light} {Pollution} {Weather}}.
\newblock
  \bibinfo{howpublished}{\url{http://dc.zah.uni-heidelberg.de/lightmeter/q/weather/form}};
  \bibinfo{year}{2009}.
\newblock \bibinfo{note}{Accessed:10 Sept 2017}.
\bibitem[{{Aceituno} et~al.(2011){Aceituno}, {S{\'a}nchez}, {Aceituno},
  {Galad{\'{\i}}-Enr{\'{\i}}quez}, {Negro}, {Soriguer} et~al.}]{aceituno2011}
\bibinfo{author}{{Aceituno}\xfnm[ J.]}, \bibinfo{author}{{S{\'a}nchez}\xfnm[
  S.F.]}, \bibinfo{author}{{Aceituno}\xfnm[ F.J.]},
  \bibinfo{author}{{Galad{\'{\i}}-Enr{\'{\i}}quez}\xfnm[ D.]},
  \bibinfo{author}{{Negro}\xfnm[ J.J.]}, \bibinfo{author}{{Soriguer}\xfnm[
  R.C.]}, et~al.
\newblock \bibinfo{title}{An {All}-{Sky} {Transmissio}{ Monitor}:{ASTMON}}.
\newblock \bibinfo{journal}{Publ Astron Soc Pac}
  \bibinfo{year}{2011};\bibinfo{volume}{123}:\bibinfo{pages}{1076--1086}.
\newblock \DOIprefix\doi{10.1086/661918}.
\bibitem[{{Falchi}(2011)}]{falchi2011}
\bibinfo{author}{{Falchi}\xfnm[ F.]}.
\newblock \bibinfo{title}{Campaign of sky brigthness and extinction
  measurements using a portable {CCD} camera}.
\newblock \bibinfo{journal}{Monthly Notices of the Royal Astronomical Society}
  \bibinfo{year}{2011};\bibinfo{volume}{412}(\bibinfo{number}{1}):\bibinfo{pages}{33--48}.
\bibitem[{{Nievas}(2013)}]{nievas2003}
\bibinfo{author}{{Nievas}\xfnm[ M.]}.
\newblock \bibinfo{title}{Absolute photometry and {Sky} {Brightness} with
  {ASTMON}-{UCM}}.
\newblock Master's thesis; Universidad Complutense de Madrid;
  \bibinfo{year}{2013}.
\bibitem[{{Zamorano} et~al.(2015){Zamorano}, {Nievas}, {S{\'a}nchez de Miguel},
  {Tapia}, {Garc{\'{\i}}a}, {Pascual} et~al.}]{2015IAUGA}
\bibinfo{author}{{Zamorano}\xfnm[ J.]}, \bibinfo{author}{{Nievas}\xfnm[ M.]},
  \bibinfo{author}{{S{\'a}nchez de Miguel}\xfnm[ A.]},
  \bibinfo{author}{{Tapia}\xfnm[ C.]}, \bibinfo{author}{{Garc{\'{\i}}a}\xfnm[
  C.]}, \bibinfo{author}{{Pascual}\xfnm[ S.]}, et~al.
\newblock \bibinfo{title}{Low-cost photometers and open source software for
  {Light} {Pollution} research}.
\newblock \bibinfo{journal}{IAU General Assembly}
  \bibinfo{year}{2015};\bibinfo{volume}{22}:\bibinfo{eid}{2254626}.
\bibitem[{Bar\'{a} et~al.(2014)Bar\'{a}, Nievas, de~Miguel and
  Zamorano}]{bara14}
\bibinfo{author}{Bar\'{a}\xfnm[ S.]}, \bibinfo{author}{Nievas\xfnm[ M.]},
  \bibinfo{author}{de~Miguel\xfnm[ A.S.]}, \bibinfo{author}{Zamorano\xfnm[
  J.]}.
\newblock \bibinfo{title}{Zernike analysis of all-sky night brightness maps}.
\newblock \bibinfo{journal}{Applied Optics}
  \bibinfo{year}{2014};\bibinfo{volume}{53}(\bibinfo{number}{12}):\bibinfo{pages}{2677--2686}.
\newblock \DOIprefix\doi{10.1364/AO.53.002677}.
\bibitem[{Ribas(2013)}]{ribas2013}
\bibinfo{author}{Ribas S. J. Canal-Domingo\xfnm[ R.]}.
\newblock \bibinfo{title}{{Measuring} {Light} {Pollution} in {Montsec}: a
  protected area}.
\newblock \bibinfo{journal}{Proceedings of the International Conference on
  Light Pollution: Theory, Modelling and Mesurements, Smolenice(Slovak
  Republic} \bibinfo{year}{2013};:\bibinfo{pages}{113--117}.
\bibitem[{ram(2011)}]{ramirez2011}
\bibinfo{title}{{Ramirez} {Moreta, P.,} {Brillo} de fondo de cielo con
  astmon-ucm, {PhD} thesis}.
\newblock \bibinfo{howpublished}{\url{http://eprints.ucm.es/15000/}};
  \bibinfo{year}{2011}.
\bibitem[{{Jechow} et~al.(2017){Jechow}, {Koll{\'a}th}, {Ribas}, {Spoelstra},
  {H{\"o}lker} and {Kyba}}]{jechow2017a}
\bibinfo{author}{{Jechow}\xfnm[ A.]}, \bibinfo{author}{{Koll{\'a}th}\xfnm[
  Z.]}, \bibinfo{author}{{Ribas}\xfnm[ S.J.]},
  \bibinfo{author}{{Spoelstra}\xfnm[ H.]}, \bibinfo{author}{{H{\"o}lker}\xfnm[
  F.]}, \bibinfo{author}{{Kyba}\xfnm[ C.]}.
\newblock \bibinfo{title}{Imaging and mapping the impact of clouds on skyglow
  with all-sky photometry}.
\newblock \bibinfo{journal}{Scientific Reports}
  \bibinfo{year}{2017};\bibinfo{volume}{7}:\bibinfo{pages}{6741}.
\bibitem[{Jechow et~al.(2017)Jechow, Koll{\'a}th, Lerner, H{\"a}nel, Shashar,
  H{\"o}lker et~al.}]{jechow2017b}
\bibinfo{author}{Jechow\xfnm[ A.]}, \bibinfo{author}{Koll{\'a}th\xfnm[ Z.]},
  \bibinfo{author}{Lerner\xfnm[ A.]}, \bibinfo{author}{H{\"a}nel\xfnm[ A.]},
  \bibinfo{author}{Shashar\xfnm[ N.]}, \bibinfo{author}{H{\"o}lker\xfnm[ F.]},
  et~al.
\newblock \bibinfo{title}{Measuring light pollution with fisheye lens imagery
  from a moving boat, a proof of concept}.
\newblock \bibinfo{journal}{International Journal of Sustainable Lighting}
  \bibinfo{year}{2017};\bibinfo{volume}{19}(\bibinfo{number}{1}):\bibinfo{pages}{15--25}.
\bibitem[{{Zotti}(2007)}]{zotti2007}
\bibinfo{author}{{Zotti}\xfnm[ G.]}.
\newblock \bibinfo{title}{Computer graphics in historical and modern sky
  observations}.
\newblock Ph.D. thesis; Vienna University of Technology; \bibinfo{year}{2007}.
\bibitem[{{Kocifaj} et~al.(2015){Kocifaj}, {Solano Lamphar} and
  {Kundracik}}]{kocifaj+2015}
\bibinfo{author}{{Kocifaj}\xfnm[ M.]}, \bibinfo{author}{{Solano Lamphar}\xfnm[
  H.]}, \bibinfo{author}{{Kundracik}\xfnm[ F.]}.
\newblock \bibinfo{title}{Retrieval of garstang's emission function from
  all-sky camera images}.
\newblock \bibinfo{journal}{Monthly Notices of the Royal Astronomical Society}
  \bibinfo{year}{2015};\bibinfo{volume}{453}(\bibinfo{number}{1}):\bibinfo{pages}{819--827}.
\bibitem[{{Solano Lamphar} and {Kundracik}(2014)}]{lamphar2014}
\bibinfo{author}{{Solano Lamphar}\xfnm[ H.]},
  \bibinfo{author}{{Kundracik}\xfnm[ F.]}.
\newblock \bibinfo{title}{A microcontroller-based system for automated and
  continuous sky glow measurements with the use of digital single-lens reflex
  cameras}.
\newblock \bibinfo{journal}{Lighting Research \& Technology}
  \bibinfo{year}{2014};\bibinfo{volume}{46}(\bibinfo{number}{1}):\bibinfo{pages}{20--30}.
\bibitem[{{Kloppenborg} et~al.(2013){Kloppenborg}, {Pieri}, {Eggenstein},
  {Maravelias} and {Pearson}}]{kloppenborg2013}
\bibinfo{author}{{Kloppenborg}\xfnm[ B.K.]}, \bibinfo{author}{{Pieri}\xfnm[
  R.]}, \bibinfo{author}{{Eggenstein}\xfnm[ H.B.]},
  \bibinfo{author}{{Maravelias}\xfnm[ G.]}, \bibinfo{author}{{Pearson}\xfnm[
  T.]}.
\newblock \bibinfo{title}{A demonstration of accurate wide-field {V}-band
  photometry using a consumer-grade {DSLR} camera}.
\newblock \bibinfo{journal}{arXiv preprint arXiv:13036870}
  \bibinfo{year}{2013};.
\bibitem[{Mumpuni et~al.(2010)Mumpuni, Kesumaningrum and Muhamad}]{mumpuni2009}
\bibinfo{author}{Mumpuni\xfnm[ E.S.]}, \bibinfo{author}{Kesumaningrum\xfnm[
  R.]}, \bibinfo{author}{Muhamad\xfnm[ J.]}.
\newblock \bibinfo{title}{{Simple} {Photometry} to {Measure} the {Sky}
  {Brightness} using a {DSLR} {Camera}}.
\newblock \bibinfo{journal}{Proceedings of the Conference of the Indonesian
  Society of Astronomy and Astrophysics}
  \bibinfo{year}{2010};\bibinfo{volume}{29}:\bibinfo{pages}{155--158}.
\bibitem[{Hiscocks and Eng(2011)}]{hiscocks2013}
\bibinfo{author}{Hiscocks\xfnm[ P.D.]}, \bibinfo{author}{Eng\xfnm[ P.]}.
\newblock \bibinfo{title}{Measuring {Luminance} with a digital camera}.
\newblock \bibinfo{journal}{Syscomp Electronic Design Limited}
  \bibinfo{year}{2011};.
\bibitem[{{H\"anel}(2015)}]{haenel2015}
\bibinfo{author}{{H\"anel}\xfnm[ A.]}.
\newblock \bibinfo{title}{Eichung von {Fischaugenaufnahmen} mit dem {Sky}
  {Quality} {Meter}}.
\newblock \bibinfo{journal}{Journal für Astronomie, Vereinigung der
  Sternfreunde}
  \bibinfo{year}{2015};\bibinfo{volume}{54}(\bibinfo{number}{3}):\bibinfo{pages}{33--36}.
\bibitem[{Duriscoe et~al.(2007)Duriscoe, Luginbuhl and
  Moore}]{1538-3873-119-852-192}
\bibinfo{author}{Duriscoe\xfnm[ D.M.]}, \bibinfo{author}{Luginbuhl\xfnm[
  C.B.]}, \bibinfo{author}{Moore\xfnm[ C.A.]}.
\newblock \bibinfo{title}{Measuring {Night}-{Sky} {Brightness} with a
  {Wide}-{Field} {CCD} {Camera}}.
\newblock \bibinfo{journal}{Publications of the Astronomical Society of the
  Pacific}
  \bibinfo{year}{2007};\bibinfo{volume}{119}(\bibinfo{number}{852}):\bibinfo{pages}{192}.
\newblock \URLprefix \url{http://stacks.iop.org/1538-3873/119/i=852/a=192}.
\bibitem[{Duriscoe(2016)}]{2016duriscoe}
\bibinfo{author}{Duriscoe\xfnm[ D.M.]}.
\newblock \bibinfo{title}{Photometric indicators of visual night sky quality
  derived from all-sky brightness maps}.
\newblock \bibinfo{journal}{J Quant Spectrosc Ra}
  \bibinfo{year}{2016};\bibinfo{volume}{181}:\bibinfo{pages}{33--45}.
\bibitem[{{Cinzano} and {Falchi}(2003)}]{2003MmSAI..74..458C}
\bibinfo{author}{{Cinzano}\xfnm[ P.]}, \bibinfo{author}{{Falchi}\xfnm[ F.]}.
\newblock \bibinfo{title}{A portable wide-field instrument for mapping night
  sky brightness automatically}.
\newblock \bibinfo{journal}{Mem S A It}
  \bibinfo{year}{2003};\bibinfo{volume}{74}:\bibinfo{pages}{458}.
\bibitem[{{Falchi}(1998)}]{falchi1998}
\bibinfo{author}{{Falchi}\xfnm[ F.]}.
\newblock \bibinfo{title}{Luminanza artificiale del cielo notturno in
  {Italia}}.
\newblock Master's thesis; Universit{\`a} di Milano; \bibinfo{year}{1998}.
\bibitem[{{Benn} and {Ellison}(2007)}]{benn2007}
\bibinfo{author}{{Benn}\xfnm[ C.]}, \bibinfo{author}{{Ellison}\xfnm[ S.]}.
\newblock \bibinfo{title}{La {Palma} technical note 115}.
\newblock \bibinfo{journal}{Roque de los Muchachos Observatory}
  \bibinfo{year}{2007};.
\bibitem[{Puschnig et~al.(2014)Puschnig, Posch and
  Uttenthaler}]{puschnig2014night}
\bibinfo{author}{Puschnig\xfnm[ J.]}, \bibinfo{author}{Posch\xfnm[ T.]},
  \bibinfo{author}{Uttenthaler\xfnm[ S.]}.
\newblock \bibinfo{title}{Night sky photometry and spectroscopy performed at
  the {Vienna} {University} {Observatory}}.
\newblock \bibinfo{journal}{J Quant Spectrosc Ra}
  \bibinfo{year}{2014};\bibinfo{volume}{139}:\bibinfo{pages}{64--75}.
\bibitem[{Marin(2009)}]{Aube2007}
\bibinfo{author}{Marin\xfnm[ C.]}.
\newblock \bibinfo{title}{Starlight: a common heritage}.
\newblock \bibinfo{journal}{Proceedings of the International Astronomical
  Union}
  \bibinfo{year}{2009};\bibinfo{volume}{5}(\bibinfo{number}{S260}):\bibinfo{pages}{449--456}.
\bibitem[{Consortium et~al.(2010)}]{tcta2010design}
\bibinfo{author}{Consortium\xfnm[ T.]}, et~al.
\newblock \bibinfo{title}{{Design} {Concepts} for the {Cherenkov} {Telescope}
  {Array}}.
\newblock \bibinfo{journal}{Experimental Astronomy}
  \bibinfo{year}{2010};\bibinfo{volume}{32}(\bibinfo{number}{3}).
\bibitem[{van Grunsven et~al.(2014)van Grunsven, Donners, Boekee, Tichelaar,
  Van~Geffen, Groenendijk et~al.}]{van2014spectral}
\bibinfo{author}{van Grunsven\xfnm[ R.H.]}, \bibinfo{author}{Donners\xfnm[
  M.]}, \bibinfo{author}{Boekee\xfnm[ K.]}, \bibinfo{author}{Tichelaar\xfnm[
  I.]}, \bibinfo{author}{Van~Geffen\xfnm[ K.]},
  \bibinfo{author}{Groenendijk\xfnm[ D.]}, et~al.
\newblock \bibinfo{title}{Spectral composition of light sources and insect
  phototaxis, with an evaluation of existing spectral response models}.
\newblock \bibinfo{journal}{Journal of insect conservation}
  \bibinfo{year}{2014};\bibinfo{volume}{18}(\bibinfo{number}{2}):\bibinfo{pages}{225--231}.
\bibitem[{tec(2016)}]{techno}
\bibinfo{title}{Technoteam}.
\newblock \bibinfo{howpublished}{\url{http://www.technoteam.de}};
  \bibinfo{year}{2014-2016}.
\newblock \bibinfo{note}{Accessed: 30 Jan 2017}.
\bibitem[{Schmidt(2011)}]{schmidt2011}
\bibinfo{author}{Schmidt\xfnm[ W.]}.
\newblock \bibinfo{title}{Lichtonderzoek zuidholland}.
\newblock \bibinfo{journal}{Sotto le Stelle, 4/2011}
  \bibinfo{year}{2011};\URLprefix
  \url{http://www.sotto.nl/rapporten/Eindrapport%20Zuid-Holland.pdf};
  \bibinfo{note}{accessed: 11 Sept 2017}.

\end{thebibliography}
\end{document}